\documentclass[letterpaper,twocolumn,10pt]{article}
\usepackage{usenix}
\usepackage{tikz}
\usepackage{amsmath}

\usepackage{pifont}
\usepackage{svg}
\usepackage{filecontents}
\usepackage{circledsteps}
\usepackage{xcolor}
\pgfkeys{/csteps/inner ysep=3pt}
\pgfkeys{/csteps/inner xsep=4pt}
\newcommand{\blackcircled}[1]{%
  \tikz[baseline=(char.base)] \node[fill=black, text=white, circle, inner sep=0.5mm] (char) {#1};}

\usepackage{caption}
\usepackage{subcaption}
\usepackage{listings}
\usepackage{ltablex}
\usepackage{multirow}
\usepackage{graphicx}
\usepackage{adjustbox}
\usepackage[normalem]{ulem}
\usepackage{rotating}
\usepackage{colortbl}
\usepackage{hhline}
\usepackage[framemethod=TikZ]{mdframed}
\usepackage{tabularray}
\usepackage{enumitem}
\usepackage{dblfloatfix}
\usepackage{placeins}
\usepackage{float}
\usepackage[affil-it]{authblk}

\newcommand{\baseNumBin}{33}
\newcommand{\numbin}{61}
\newcommand{\numuniquebugs}{187}
\newcommand{\numoracles}{313}
\newcommand{\numfuzz}{9}
\newcommand{\cpuyears}{10}
\newcommand{\codeangle}[1]{\textcolor{purple}{\texttt{<#1>}}}
\newcommand{\uniqueCWECount}{34}
\newcommand{\name}{\textsc{FirmReBugger}}
\newcommand{\namebench}{\textsc{FirmBench}}
\newcommand{\namebenchdma}{\textsc{FirmBenchDMA}}
\newcommand{\dyma}{\textsc{DyMA-Fuzz}}
\newcommand{\namebenchX}{\textsc{FirmBenchX}}
\newcommand{\multifuzz}[0]{\textsc{MultiFuzz}}
\newcommand{\dice}[0]{\textsc{DICE}}
\newcommand{\fuzzware}[0]{\textsc{Fuzzware}}

\newcommand{\ptwoim}[0]{{\sc P$^2$IM}}
\newcommand{\uemu}[0]{{\sc $\mu$Emu}} 
\newcommand{\ember}[0]{{\sc Ember-IO}}
\newcommand{\fuzzwareicicle}[0]{{\sc Fuzzware-Icicle}}
\newcommand{\splits}[0]{{\sc SplITS}}
\newcommand{\semu}[0]{{\sc SEmu}}
\newcommand{\hoedur}[0]{{\sc Hoedur}}
\newcommand{\gdma}[0]{{\sc GDMA}}
\newcommand{\bin}[1]{\textsf{#1}}
\newcommand{\missing}{\makebox[2em][c]{--}}

\usepackage{tikz}

\definecolor{codegreen}{rgb}{0,0.6,0}
\definecolor{codegray}{rgb}{0.5,0.5,0.5}
\definecolor{codepurple}{rgb}{0.58,0,0.82}
\definecolor{backcolour}{rgb}{0.95,0.95,0.92}
\definecolor{lightgray}{gray}{0.95}
\newcommand{\tick}{\ding{51}}
\newcommand{\cross}{\ding{55}}
\usepackage{comment}
\usepackage[frozencache=true,cachedir=minted-cache]{minted}
\AtBeginEnvironment{minted}{
  }
\setminted[cpp]{
    bgcolor=lightgray
}

\hypersetup{
pdfauthor={}, 
pdfsubject={}, 
pdfkeywords={}, 
pdfcreator={} 
}

\begin{document}
\date{}

\title{\Large \bf FirmReBugger: A Benchmark Framework for Monolithic Firmware Fuzzers}

\author[1]{Mathew Duong}
\author[1]{Michael Chesser}
\author[1]{Guy Farrelly}
\author[2]{Surya Nepal}
\author[1]{Damith C. Ranasinghe}

\affil[1]{University of Adelaide}
\affil[2]{Data61 CSIRO}

\maketitle

\begin{abstract}
Monolithic Firmware is widespread. Unsurprisingly, fuzz testing firmware is an active research field with new advances addressing the \textit{unique} challenges in the domain. However, understanding and evaluating improvements by deriving metrics such as code coverage and unique crashes are problematic, leading to a desire for a reliable bug-based benchmark. To address the need, we design and build \name{}, a holistic framework for fairly assessing monolithic firmware fuzzers with a realistic, diverse, bug-based benchmark.

\name{} proposes using bug oracles---C syntax expressions of bug descriptors---with an \textit{interpreter} to automate analysis and \textit{accurately} report on bugs discovered, discriminating between states of \textit{detected}, \textit{triggered}, \textit{reached} and \textit{not reached}.
Importantly, our idea  of benchmarking does not modify the target binary and simply \textit{replays} fuzzing seeds to \textit{isolate} the benchmark implementation from the fuzzer while providing a simple means to extend with new bug oracles.

Further, analyzing fuzzing roadblocks, we created \namebench{}, a set of diverse, real-world binary targets with  \numoracles{} software bug oracles. Incorporating our analysis of roadblocks challenging monolithic firmware fuzzing, the bench provides for rapid evaluation of future advances. We implement \name{} in a \textit{FuzzBench-for-Firmware} type service and use \namebench{} to evaluate \numfuzz{} state-of-the art monolithic firmware fuzzers in the style of a reproducibility study, using a 10 CPU-year effort, to report our findings.
\end{abstract}

\section{Introduction}
The surge in embedded systems, from cube-satellites to smart door locks and autonomous vehicles, is driven by small, low cost, and low-power microcontrollers. The proliferation of embedded systems creates more targets, new attack vectors, opportunities, and incentives for adversaries~\cite{papp2015embedded}. Testing firmware to identify vulnerabilities prior to exploitation or device failure is critically important. For example, Hanna \textit{et al.}'s~\cite{hanna2011take} assessment of defibrillator firmware in a commercial product revealed critical vulnerabilities, including a buffer overflow that could allow remote code execution. 

Unsurprisingly, embedded systems' rising prevalence and associated security concerns are driving  advances in automated software testing methods like fuzzing. However, uncovering potential security vulnerabilities in firmware with fuzzing is challenged by hardware limitations, resource constraints, and the complexity of direct firmware interactions with hardware. Consequently, firmware fuzzing is an active topic of  research interest where recent methods~\cite{feng2020p2im,uEmu,scharnowski2022fuzzware,zhou2022semu,dice2021,farrelly2023ember,farrelly2023splits,chesser2023icicle,chessermultifuzz} re-host firmware using emulation on more resourceful computing platforms to bypass the resource limitations of embedded hardware. Emulation provides the ability to add sophisticated instrumentation to guide a fuzzer, facilitates introspection, and enables better detection and reproducibility of crashes whilst avoiding the need to synchronize hardware and emulation environments with \textit{device-in-the-loop} methods. Despite the benefits of re-hosting, applying automated testing methods like fuzzing  remains challenging, especially for monolithic firmware. 

Monolithic firmware directly manages all interactions with hardware, operating without support or supervisory control of an operating system, and combines all essential software components into a single unified entity. Most notably, in contrast to traditional applications, monolithic firmware interacts directly with peripherals (such as modems, GPS units, and sensors) through Memory Mapped Input Output (MMIO) or Direct Memory Access (DMA) mechanisms and interrupts. These low-level interactions are complex and may be across a diverse group of peripherals with behaviors differing between manufacturers. This poses significant challenges for employing a de-facto, grey-box, mutation-based fuzzing method to drive re-hosted firmware to discover software bugs.

Advances in fuzz testing firmware aim to overcome the pertinent challenges to achieve scalable, efficient, and effective fuzzers by facilitating the firmware's interaction with a diverse group of peripheral devices across various different microcontroller designs employing a range of different instruction set architectures (ISAs). But, a key question with regards to the evaluation of fuzzers, to support both the growth and rapid developments of new techniques is:
\begin{mdframed}[backgroundcolor=black!10,rightline=false,leftline=false,topline=false,bottomline=false,roundcorner=2mm] 
    How do we fairly assess improvements to the \textit{bug finding} capabilities of monolithic firmware fuzzers?
\end{mdframed}

\subsection{Unique Problems Challenging Assessment}
Assessing the performance of fuzzers, in general, relies on \textit{empirical} means to evaluate their bug-discovering capabilities. Code coverage and unique crash counts are adopted from domains outside of firmware targets to compare if one firmware fuzzer is ``better'' than another~\cite{feng2020p2im,uEmu,scharnowski2022fuzzware,zhou2022semu,farrelly2023ember,farrelly2023splits,chesser2023icicle,chessermultifuzz, scharnowskigdma}.  
Unfortunately, adopting these accepted performance measures to evaluate monolithic firmware fuzzers is not without its unique set of problems. We delve into these issues in detail in Section~\ref{sec:challenges in fuzzing} and summarised them below:   

\begin{itemize}[itemsep=2pt,parsep=1pt,topsep=3pt,labelindent=5pt,leftmargin=12pt]
    \item \textit{Code Coverage}.
    The relationship of coverage achieved by a fuzzer to its bug-finding ability support it as a performance metric~\cite{gopinath2014code,inozemtseva2014coverage,kochhar2015code}. However, its use in evaluating monolithic firmware fuzzers can lead to misleading results. For example, a fuzzer exploiting a bug to reach otherwise unreachable code can inflate the coverage metric, giving the illusion of better performance---a scenario more commonly encountered in embedded systems (see Section~\ref{sec:prob-coverage}).

    \item \textit{Unique Crashes.~}Unique crashes are a more direct measure of the number of bugs found. But, implementing it accurately in the firmware domain is challenging. Differentiating and deduplicating crashes require capturing a bug’s context. Common identifiers, such as coverage profiles and stack hashes are prone to variability in program behavior, making consistent and reliable crash classification difficult. The problem is further exacerbated in the firmware domain where program flow is more complex due to the interrupt driven nature of firmware (see Section~\ref{sec:prob-unique-crashes}).
    
\end{itemize}

A bug-based benchmark is recognized as an approach to alleviate the imperfect nature of coverage and problems with unique crash metrics. In general, bug-based benchmarks such as LAVA \cite{dolan2016lava} with LAVA-M, Magma \cite{hazimeh2020magma}, UNIFUZZ \cite{li2021unifuzz}, Google's FuzzBench \cite{metzman2021fuzzbench}, FixReverter \cite{zhang2022fixreverter}, and FuzzProBench \cite{profuzzbench} are proposed for \textit{non-firmware} targets. However, a benchmark for evaluating monolithic firmware fuzzers does not yet exist. Notably, recent method for designing a benchmark in FixReverter~\cite{zhang2022fixreverter} require access to source code, which is often unavailable with firmware images. And the bug-oracle approach in Magma~\cite{hazimeh2020magma} integrates the benchmark into the fuzzing logic and leads to the reported leaky oracle problem~\cite{hazimeh2020magma}. In addition, given the developing nature of firmware fuzzing there is a lack of effective tooling for automating the triaging of bugs. The alternative---curating direct performance results such as time-to-find bugs---necessitates manual crash analysis and triaging for each fuzzer, which is both time-consuming and error-prone. 

Consequently, we are motivated to design a bug-based benchmarking framework for monolithic firmware fuzzers. We consider the unique challenges posed in the domain and facilitate automated and fair evaluation of fuzzers. We summarize our efforts in the following sections.
 
\subsection{Our Work}
\label{sec:OurWork}

We analyze the problems with empirical performance evaluation for firmware images (Section~\ref{sec:challenges in fuzzing}) and propose \name{}, a new benchmark framework together with \namebench{} target sets---challenging, real-world, benchmark binaries with a diverse set of bugs. Importantly, we curate binary targets through a careful consideration of existing monolithic firmware fuzzing challenges we discuss in Section~\ref{Common Roadblocks in Fuzzing Firmware}.

 \vspace{2mm}
\noindent\textbf{Benchmark Design Goals.~}
First, we extend the collective wisdom from the design of benchmarks outside of monolithic firmware and curate a desirable set of goals for a firmware benchmark. Notably, recent fuzzing benchmark studies, such as those highlighted by Zhang et al. \cite{zhang2022fixreverter} and Hazimeh et al. \cite{hazimeh2020magma}, have identified necessary objectives for creating an \textit{effective} and \textit{reliable} fuzzing benchmark. Drawing inspiration from these insights, we present the following goals we deem essential in developing a comprehensive benchmarking framework. The benchmark:

\begin{description}[itemsep=2pt,parsep=1pt,topsep=3pt,labelindent=6pt,leftmargin=23pt]
 \item[\textit{G1}]Should not affect the fuzzing process (the leaky oracle problem).
  \item[\textit{G2}]Should provide clear indicators of bugs triggered.
  \item[\textit{G3}]Should use relevant, real-world target programs.
  \item[\textit{G4}]Should contain targets with realistic and relevant bugs.
  \item[\textit{G5}]Should defend against overfitting of fuzzers to targets.
\end{description}

To address \textit{G1}, we isolate the fuzzer from the benchmark triaging process. This ensures that the benchmark itself does not influence fuzzing outcomes, mitigating the leaky oracle problem. Specifically, \name{} uses \textit{bug oracles}. We create bug expressions to formally describe the conditions to confirm a specific bug. These expressions are interpreted by replaying both crashing and non-crash inputs (fuzzing inputs) to determine whether a specific bug is reached, triggered, or detected. By this analysis, \name{} automatically computes essential bug-based performance metrics---\textit{G2}. 

To satisfy \textit{G3} and \textit{G4}, \name{} uses a curated set of binaries covering widely used real-world firmware and relevant bugs. The binaries span multiple system libraries and hardware contexts, ensuring realistic evaluation scenarios for fuzzers. Finally \textit{G5} is addressed providing a diverse and extensible benchmark set. This prevents overfitting by exposing fuzzers to a variety of binaries, bugs, and conditions rather than letting them specialize on a narrow set of targets.

The resulting \name{} framework is supported by \numbin{} binary targets and \numoracles{} software bug oracles within three sets of compiled, ready-to-fuzz binaries  in \namebench{}, \namebenchdma{} and \namebenchX{}; with \namebenchdma{} and \namebenchX{} created to capture more complex challenges in fuzzing firmware targets.

\vspace{2mm}
\noindent\textbf{Contributions.~} We make the following contributions: 

\begin{itemize}[itemsep=2pt,parsep=1pt,topsep=3pt,labelindent=5pt,leftmargin=12pt]
    \item We propose a first fuzzing benchmark framework, \name{}, for 
    monolithic firmware fuzzers. Importantly, our approach addresses the problem of leaky oracles, defending against overfitting in benchmarks, and provides ground-truth bug-based metrics.
    \item Our key idea is to exploit the introspection capabilities of emulators to automatically translate virtual CPU state to bugs states using simple-to-craft expressions of bug descriptions (bug oracles) during replay of fuzzing seeds.  
    \item We implement and open-source \name{} as a \textit{FuzzBench-for-Firmware} type service.
    \item We curate a benchmark set of \numbin{} diverse binary targets with \numoracles{} bug oracles, across three datasets, incorporating fuzzing challenges from our analysis to facilitate fair assessment of current and future advances in firmware fuzzing.

\end{itemize}

\vspace{1mm}
\begin{mdframed}[backgroundcolor=blue!10,rightline=false,leftline=false,topline=false,bottomline=false,roundcorner=2mm] 

Importantly, we use \name{} to conduct an extensive benchmark study on \numfuzz{} state-of-the-art firmware fuzzers~\cite{farrelly2023ember,scharnowski2022fuzzware,chesser2023icicle,zhou2022semu,dice2021, farrelly2023splits, chessermultifuzz, scharnowski2023hoedur, scharnowskigdma}, across \numbin{} binaries with \numuniquebugs{} bugs. Each fuzzer underwent 10 repeated 24-hour fuzzing runs, culminating in a total evaluation effort equivalent to \cpuyears{} CPU-years.
\end{mdframed}

\vspace{2mm}
\noindent\textbf{Paper Organisation.~}
We re-visit approaches to monolithic firmware fuzzing in Section~\ref{sec:background}. Section~\ref{sec:challenges in fuzzing}  investigates \textit{problems} with current empirical performance measures. Our bug-based benchmark framework, \name{}, and its implementation is in Section~\ref{sec:firmrebugger}. Importantly, in Section~\ref{sec:BechnmarkTargets} we investigate current \textit{firmware fuzzing challenges facing re-hosting frameworks} to construct a diverse set of targets incorporating the challenges to provide for future advances in firmware fuzzing. Then, in Section~\ref{sec:experiments}, we present and analyze the results of our own extensive benchmark study of state-of-the-art fuzzers.

\section{Background on Firmware Fuzzing}
\label{sec:background}

\begin{figure*}[t]
    \centering
    \includegraphics[width=\textwidth]{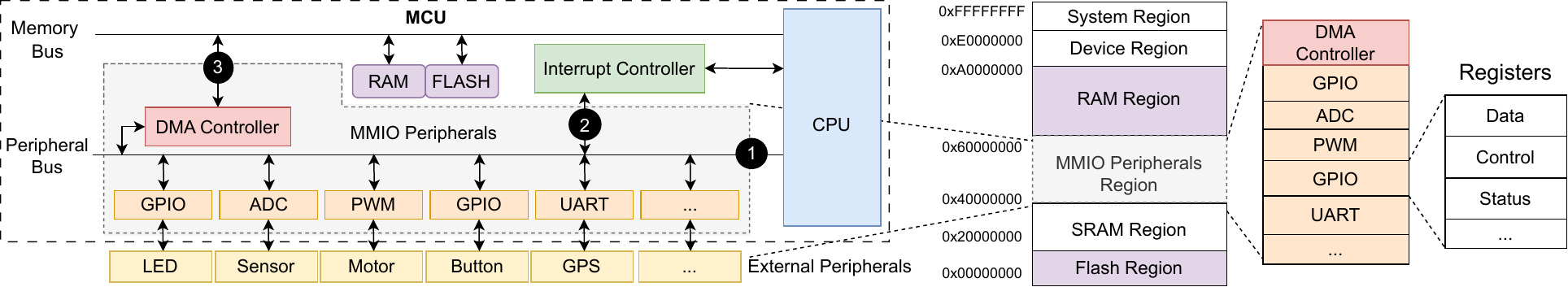}
    \caption{The internal architecture of a microcontroller, illustrating the firmware interactions with peripheral devices.}
    \label{figure:MCU-overview}
\end{figure*}

\textbf{Monolithic Firmware.~}Firmware, software for the hardware, is found in embedded systems across industries, ranging from automotive systems, industrial control systems, medical devices, to consumer electronics. Monolithic firmware targets microcontroller units (MCUs). These low-cost MCUs feature fewer resources than desktop processors and use simpler instruction set architectures (ISAs) such as ARM or RISC-V.
MCUs incorporate a range of different peripherals such as GPIO, ADC, UART or SPI from a diverse set as illustrated in Figure~\ref{figure:MCU-overview}. These peripherals are used for real-time interaction with sensors and actuators, allowing complex products such as CNCs~\cite{grbl} or drones~\cite{betaflight} to be built. Monolithic firmware typically lacks the abstraction provided by an operating system, instead managing all critical tasks internally, including hardware initialization, peripheral communication, and event handling. This design tightly couples the firmware with the microcontroller unit (MCU) and its peripherals. Several interfaces are used to interact with peripherals:

\blackcircled{1} \textbf{Memory Mapped Input Output (MMIO)} is a mechanism for allocating a dedicated region of the system's memory address space to the registers of peripheral devices. 
As shown in Figure~\ref{figure:MCU-overview}, within the MMIO Peripherals Region, each peripheral device contains multiple memory-mapped registers---control, status, and data registers---that provide direct access and control over the device's state, configuration and data. 

\blackcircled{2} \textbf{Direct Memory Access (DMA)} facilitates efficient data transfer between peripherals and system memory with minimal CPU involvement. The firmware configures the DMA controller by specifying memory buffers
and a transfer size.
Once initiated, the DMA controller autonomously moves data between the peripheral and memory. When the transfer is completed, the firmware is notified via an interrupt, after which the CPU can directly process the data in memory.

\blackcircled{3} \textbf{Interrupts} are signals generated by hardware to notify the firmware of events 
that require immediate attention. When an interrupt is triggered, the CPU jumps to the associated interrupt handler and processes the signal, before returning to the prior point in execution.

\vspace{1mm}
\noindent\textbf{Firmware Fuzzing.~}
Hardware limitations, resource constraints, and the complexity of embedded system interactions make fuzzing firmware challenging. Recent work has explored approaches to address the complexities of employing fuzzing for testing firmware security. Broadly, these approaches can be split into: i)~Hardware-based approaches (on-device fuzzing and hardware-in-the-loop methods)~\cite{corteggiani2018inception,eisele2023gdb-fuzz,zaddach2014avatar,muench2018avatar}; and ii)~emulation-based approaches.

On-device fuzzing, while ideal for execution accuracy, is often impractical due to the limited computational power of embedded devices, which results in poor fuzzing throughput~\cite{feng2020p2im}. Similarly, hardware-in-the-loop fuzzing introduces additional inefficiencies, as synchronization between the physical hardware and the emulated environment incurs substantial overhead~\cite{scharnowski2022fuzzware}. Moreover, both approaches suffer from limited scalability, since they require access to specific hardware devices, making it difficult to parallelize or scale experiments across large testbeds. Consequently, researchers have proposed emulation-based approaches to support cross-architecture fuzzing, enabling faster execution speeds and greater control over the fuzzing process. 

Fuzzing firmware in an emulation environment poses a number of significant challenges.
Successful firmware emulation and fuzzing relies upon: i)~accurately handling three critical types of interactions between firmware and hardware: MMIO, DMA and interrupts; and ii)~a fuzzer capable of generating effective fuzz tests to firmware code given the diverse set of peripherals re-hosted code execution can interact with. We assess the effectiveness of solutions to these problems using empirical evaluations, next we investigate challenges faced with empirical performance evaluation of firmware fuzzers.

\section{Performance Evaluation Challenges}
\label{sec:challenges in fuzzing}
The performance of fuzzing techniques, including those applied to firmware, is evaluated through \textit{empirical} experiments. 
Various performance measures, such as code coverage and unique crash counts, are used in the literature to evaluate advances in firmware fuzzing \cite{scharnowski2022fuzzware,feng2020p2im,zhou2022semu,farrelly2023ember,chesser2023icicle}. 
However, we show that applying such metrics to evaluate firmware fuzzers is fraught with problems. In the following section, we discuss the unique difficulties faced in using established performance measures for monolithic firmware. 

\subsection{Coverage}\label{sec:prob-coverage}
Code coverage is a fundamental metric widely employed by firmware fuzzers~\cite{scharnowski2022fuzzware,farrelly2023ember,feng2020p2im, chessermultifuzz} to assess fuzzing effectiveness. In general, higher code coverage indicates greater exploration of the program's functionality, demonstrating an increasing likelihood of discovering bugs. Thus acting as a proxy for bug discovering ability. Numerous studies have explored the relationship between code coverage and bug discovery~\cite{gopinath2014code,kochhar2015code,inozemtseva2014coverage} supporting this assertion. However, the recent study by Bohme et al.~\cite{bohme2022reliability} suggested that while there is a notable \textit{correlation} between code coverage and bug discovery, there is no unanimous agreement on whether a fuzzer's superiority is solely determinable by code coverage. 

\begin{figure}[!b]
    \centering
    \includegraphics[width=0.9\linewidth]{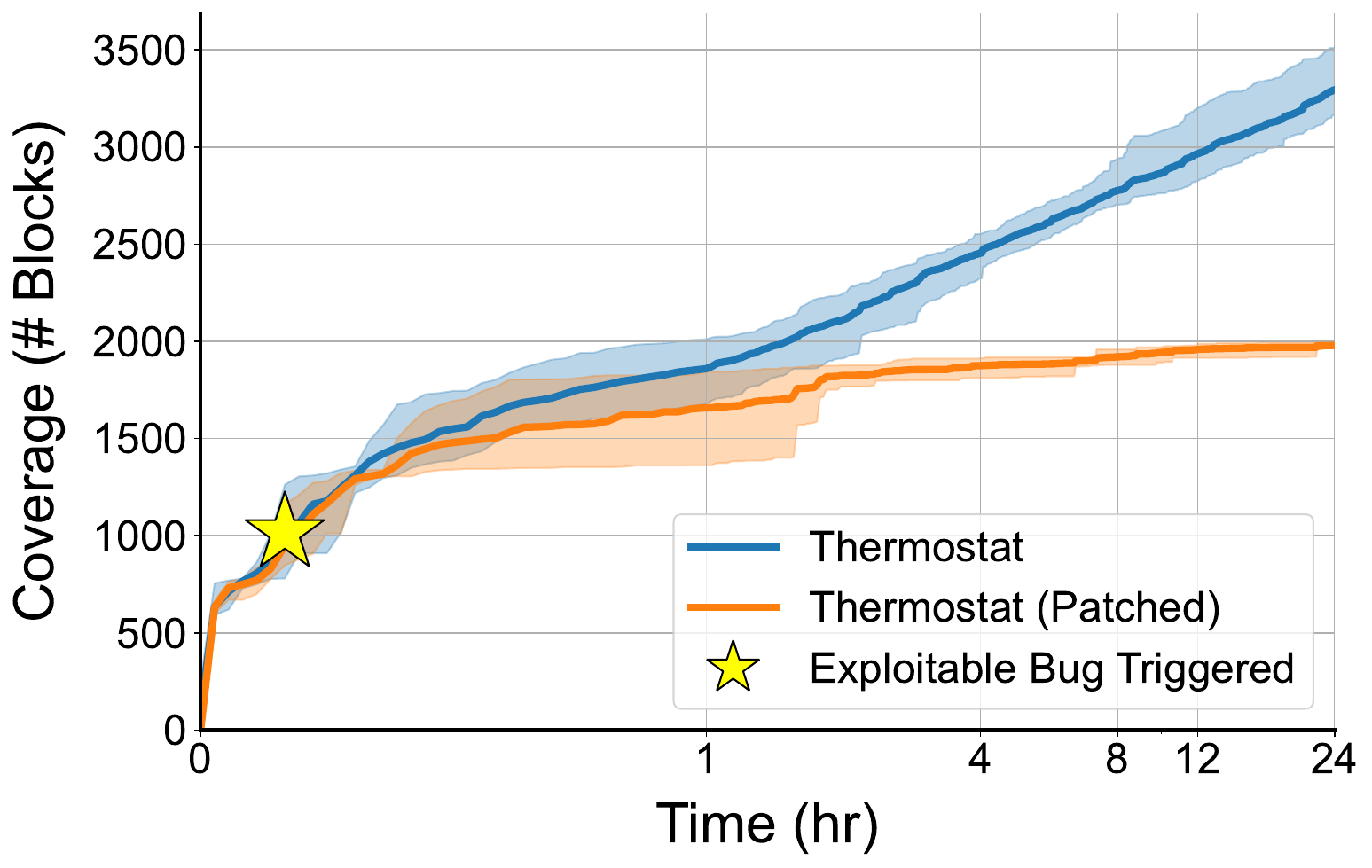}
    \caption{Number of blocks covered by \fuzzware{}~\cite{scharnowski2022fuzzware} on the \bin{Thermostat} (introduced in \textsc{Pretender}~\cite{gustafson2019Pretender}) vs. the \bin{Thermostat (patched)} binary (where the exploitable bug is removed) across five 24-hour fuzzing trials. The star denotes the mean time to trigger the exploitable bug.}
    \label{figure:bug-exploit}
\end{figure}

Bug exploitation is a problem that can hinder the collection of accurate coverage information.
On conventional targets, mechanisms such as sanitizers (e.g., ASAN~\cite{serebryany2012addresssanitizer}), control-flow integrity checks, address space layout randomization, and stronger memory isolation, typically prevent fuzz inputs from causing uncontrolled behavior. Firmware lacks these mechanisms, allowing bugs to
result in uncontrolled memory corruption. This can enable the fuzzer to manipulate the program counter by corrupting return addresses or function pointers, which can artificially inflate coverage metrics.

For example, consider the coverage plots in Figure \ref{figure:bug-exploit}. In this experiment, we used \fuzzware{}~\cite{scharnowski2022fuzzware} on the \bin{Thermostat} binary~\cite{gustafson2019Pretender} which has one known bug. At first sight, the fuzzer achieves high code coverage (blue) in Figure \ref{figure:bug-exploit}, seemingly indicating excellent performance. However, when examining the execution traces, we observe that a bug exploit allows the fuzzer to reach otherwise unreachable blocks in the code. To show the impact of the bug exploit, we patch the binary to remove the bug and fuzz the patched target. The result, orange plot in Figure~\ref{figure:bug-exploit}, shows 
more than 35\% of the original coverage resulted from exploiting the bug.

This significant reduction in coverage underscores how a bug exploit can artificially inflate code coverage.

Recent firmware fuzzers have acknowledged the problem. Both \multifuzz~\cite{chessermultifuzz} and \hoedur~\cite{scharnowski2023hoedur} highlight the impact of bug exploitation on fuzzing performance metrics. In \multifuzz~\cite{chessermultifuzz}, 6 of the 23 binaries (27\%) tested were impacted by bug exploits. The authors also mention signs of bug exploitation in 2 additional binaries during testing; however, these cases were infrequent and thus excluded from the final results. Similarly, the authors of \hoedur~\cite{scharnowski2023hoedur} note that a subset of their tested binaries 12 out of 37 (33\%) required patches to mitigate the effects of bug exploitation and preserve the integrity of code coverage as a performance measure.

\begin{figure}[b!]
    \centering
    \includegraphics[width=\columnwidth]{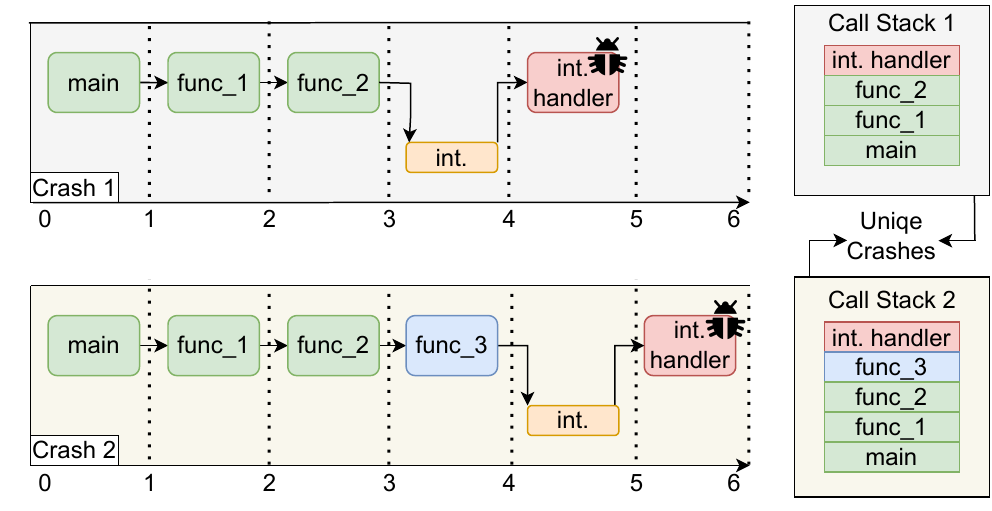}
    \caption{Illustration of two crashes with the same root cause (crash occurs in the interrupt handler). Due to the differences in interrupt timing, \underline{Crash 2} has executed an additional function (\texttt{func 3}) and as a result, has a different call stack when compared to \underline{Crash 1}. Consequently, when using the call stack to differentiate between unique crashes; \underline{Crash 1} and \underline{Crash 2} are incorrectly distinguished as unique. }
    \label{figure:stack_example}
\end{figure}

\subsection{Unique Crashes}\label{sec:prob-unique-crashes}
Counting the number of unique crashes found by each fuzzer is another commonly used metric for evaluating fuzzer effectiveness. In theory, this should be a robust measure, as it directly reflects a fuzzer's ability to discover bugs. However, in practice, relying on unique crashes as a metric is problematic for several reasons. Bugs are inherently complex, and crashes caused by the same underlying bug can manifest through different execution paths, this results in fuzzers saving many duplicate crashes.

To address this, fuzzers employ deduplication heuristics, such as crashing locations, coverage profiles, or stack hashes \cite{AFL,honggfuzz,clusterfuzz}. Despite this, studies have shown that these heuristics are inaccurate, significantly overestimating (or in some cases, underestimating) the true number of unique crashes~\cite{klees2018evaluating,manes2019art, das2020flexible, blazytko2020aurora}.

\vspace{1mm}
\noindent\textbf{Deduplication Using Common Heuristics.~}For instance, stack hash heuristics, employed in fuzzers such as Honggfuzz \cite{honggfuzz}, deduplicate crashes based on the call stack. However, with firmware, this heuristic proves less effective due to the nature of interrupts. This is illustrated in Figure \ref{figure:stack_example}, which features two crashes caused by the same root cause bug. In \underline{Crash 2}, the interrupt occurs later resulting in \underline{Crash 2} executing an additional function in \texttt{func\_3}. Which leads to a disparity in the call stacks and when relying on the call stack to deduplicate crashes this approach would erroneously mark \underline{Crash 2} as a unique crash. This effect occurs on a larger scale resulting in significantly inflated unique crash counts. Similarly, methods relying on coverage profiles, such as in  AFL~\cite{AFL}, for deduplication faces similar challenges and would struggle to deduplicate crashes for the same reasons in firmware targets.

\begin{figure}[t]
    \centering
    \includegraphics[width=\columnwidth]{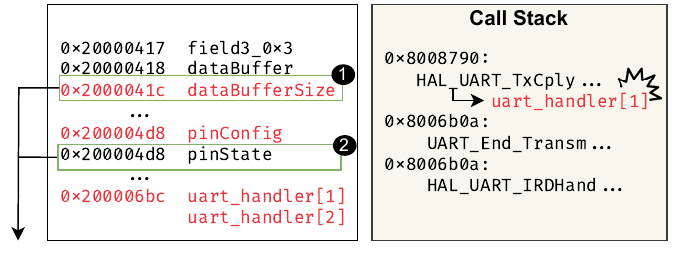}
    \caption{A simplified overview of two bugs from the \bin{Gateway} binary introduced in \ptwoim~\cite{feng2020p2im}. \texttt{dataBuffer}~\Circled{1} and \texttt{pinState}~\Circled{2} are two different vulnerable buffers defined on the heap. Due to the proximity of the buffers either buffer can overflow and corrupt the same variable---the function pointer \texttt{uart\_handler[1]}.}
    \label{figure:gateway-bug-confusion}
\end{figure}

\vspace{2mm}
Recent monolithic firmware fuzzers, such as \fuzzware{}~\cite{scharnowski2022fuzzware} use the program counter and, link register to deduplicate crashes. Although this heuristic is sufficient in many cases, it can fail in real-world targets. For example, Figure~\ref{figure:gateway-bug-confusion} shows two different buffer overflows within the \bin{Gateway} binary (introduced in \ptwoim{}~\cite{feng2020p2im}). Despite being two separate buffer overflows, the proximity of the \texttt{databuffer} and \texttt{pinState} buffers introduces a situation where both overflows can corrupt the same variable, \texttt{uart\_handler}. \textbf{Crash 1} overflows \texttt{dataBuffer}, corrupting \texttt{uart\_handler}, while \textbf{Crash 2} overflows \texttt{pinState}, corrupting of the same variable. When \texttt{uart\_handler} is later accessed in its corrupted state, both crashes will lead to a program crash at the same location. The program counter and link register-based heuristics would incorrectly categorize both cases as the same bug, effectively conflating the two bugs.

\textit{This case highlights a flaw with heuristic-based crash deduplication, where distinct bugs can be merged erroneously, leading to missed vulnerabilities.}

\subsection{Manual Crash Analysis}\label{sec:prog-manual-crash}
Due to the compounding issues we have discussed, many state-of-the-art firmware fuzzers \cite{scharnowski2022fuzzware,feng2020p2im,uEmu, farrelly2023ember, scharnowski2023hoedur} instead manually triage crashes across multiple fuzzers. However, this task is both time-consuming and error-prone. 

\begin{listing}[h]
\begin{minted}
[fontsize=\footnotesize,numbersep=4pt,xleftmargin=10pt,bgcolor=lightgray,linenos]{cpp}
//au8Buffer[2] is read from a peripheral
for (i=0; i< au8Buffer[2] /2; i++)
{
    //if au8Buffer[2] is large enough
    //a buffer overflow is triggered
    au16regs[i] = word(au8Buffer[u8byte],
                        au8Buffer[u8byte +1]);
    u8byte += 2;
}
\end{minted}
\caption{Code snippet of buffer overflow from the \bin{Heat Press} binary (introduced in~\ptwoim~\cite{feng2020p2im}).
}\label{lst:buffOverFlow}
\end{listing}

\noindent\textbf{Crash Explosion From a Single Bug.~}
This process can involve analyzing thousands of crashes resulting from a single bug to determine their uniqueness. Consider the example in Listing~\ref{lst:buffOverFlow} from the \bin{Heat Press} to illustrate a problem increasing the manual triaging task burden. In the binary, a value is read from a peripheral is stored in \texttt{au8buffer[2]} and is used as the length for the global \texttt{au16regs} buffer. If the length provided exceeds the capacity of \texttt{au16regs}, the function can corrupt many different global variables stored after \texttt{au16regs} in memory. The corruption of  different variables leads to multiple crashes in multiple locations in the code resulting in a massive number of crash reports for a single bug. 
The potential for bug exploits as described in Section~\ref{sec:prob-coverage} further complicates bug triaging and the absence of sanitizers compounds the complexity of analyzing these crashes. This increases the potential for human error and prolongs the triaging process to generate bug-based comparison results to evaluate fuzzers. Such explosion is especially common in firmware, which lacks the memory protections provided by an operating system in traditional applications.

\vspace{2mm}
\noindent\textbf{Overwhelming Effort Leads to Missing Bugs.~}
This is exemplified by the \bin{RF Door Lock} binary (\uemu{}~\cite{uEmu}), where the \fuzzware~\cite{scharnowski2022fuzzware} authors reported two distinct bugs, FW29 and FW38. In our independent evaluation~\cite{FirmReBuggerGit} of crashes resulting from \fuzzware~\cite{scharnowski2022fuzzware}, we discovered an additional bug, FRB01. We believe the miss stemmed from the overwhelming number of crashes ($\approx1600$) caused by an easily exploitable bug in the binary. The sheer volume of crashes complicates the triaging process and increases the likelihood of missing bugs. Interestingly, likely for the same reason, other fuzzers that also evaluated this binary---\splits~\cite{farrelly2023splits}, \hoedur~\cite{scharnowski2023hoedur}, \ember~\cite{farrelly2023ember}, \multifuzz~\cite{chessermultifuzz}, and \fuzzwareicicle~\cite{chesser2023icicle} overlook bug FRB01.

\begin{mdframed}[backgroundcolor=blue!10,rightline=false,leftline=false,topline=false,bottomline=false,roundcorner=2mm] 
    \textbf{Key Takeaway}: Coverage metrics are susceptible to bug exploitation artifacts. Unique crash counts are susceptible to ambiguities resulting from deduplication heuristics. The current approach of manual crash analysis across multiple fuzzers is labor intensive, error prone, and hinders progress in the field.
\end{mdframed}

\section{\name{}}\label{sec:firmrebugger}
\begin{figure}[t!]
    \centering
    \includegraphics[width=\linewidth]{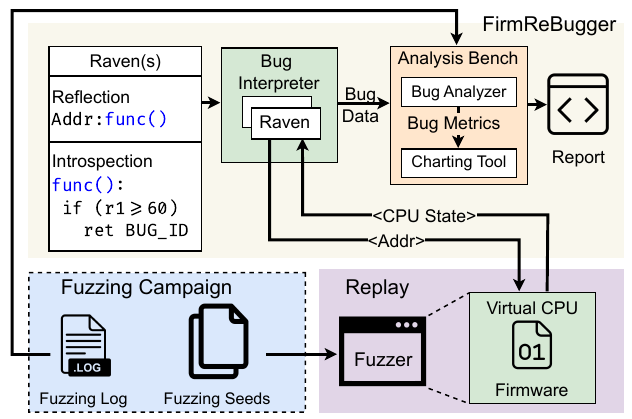}
    \caption{\name{} framework. \textit{Ravens} embed bug oracles as: i)~Reflections (virtual CPU states of interest to a bug); and ii)~Introspections (C-syntax function descriptions of state introspections to determine bug states).  The Bug Interpreter dynamically loads and translates Ravens to bug states, such as Reached \& Triggered, and unique bug IDs (Bug Data). The Bug Analyzer in the Analysis Bench processes the bug data along with crash timing data (Fuzzing Log) recorded during a fuzzing campaign into a standardized format, visualized by the Charting Tool to report performance.
    }
    \label{figure:overview}
\end{figure}

The core idea of our approach to consistently and effectively evaluating fuzzers is to automate the triaging of bugs encountered during fuzzing campaigns on curated benchmark targets using bug oracles. Our proposed benchmarking framework, achieving the aspirational goals in Section~\ref{sec:OurWork},
 to automatically curate bug-based metrics is illustrated in Figure~\ref{figure:overview}. 

\name{} exists separately from the fuzzing process.
We propose replaying both inputs and crashing seeds (Fuzzing Seeds) and automating triaging of bugs. To report on bug statistics \name{} distinguishes and determines the state of each bug encountered with respect to fuzzing inputs. Similar to benchmarks for traditional software such as FixReverter~\cite{zhang2022fixreverter} and Magma~\cite{hazimeh2020magma}, 
our approach also identifies the following states when fuzzing firmware targets:

\begin{itemize}[itemsep=2pt,parsep=1pt,topsep=3pt,labelindent=5pt,leftmargin=12pt]

\item \textbf{Not Reached.}~The input did not lead to executing the code in which the identified bug resides.  
\item \textbf{Reached.}~The input executed the lines of code in which the identified bug resides but the inputs did not meet the conditions for the bug to trigger.  
\item \textbf{Triggered.}~The input met the conditions to trigger the bug. 
\item \textbf{Detected.}~The input triggered the bug and caused a crash.
\end{itemize}
\noindent
In particular our approach:
\begin{itemize}[itemsep=2pt,parsep=1pt,topsep=3pt,labelindent=5pt,leftmargin=12pt]
    \item Isolates the fuzzer (fuzzing logic) from the automated triaging method to ensure the benchmark does not influence the fuzzing process. Effectively, \name{} replays inputs from a fuzzing campaign and utilizes the emulator to gather program state during execution and identify a bug. Importantly, the approach is agnostic to the fuzzer and benchmarking remains independent of the fuzzing logic to avoid the \textit{leaky oracle problem}---\textit{G1}.
    \item Automates analysis of fuzzing seeds, facilitating the computation of essential metrics such as time to trigger as well as discriminating between bug states (e.g. Reached)---\textit{G2}. 
    \item Curates three dedicated target subsets---\namebench{}, \namebenchdma{} and \namebenchX{}---designed to incorporate complex challenges in fuzzing real-world firmware targets to foster continuous growth in fuzzing techniques---\textit{G3} \& \textit{G4}.
    \item Defends against overfitting---where developers optimize for benchmark performance rather than general vulnerability discovery---by providing the means to add new bugs with ease. Incorporating a new bug is as simple as creating an easy-to-craft bug descriptor or a \textit{Raven}---\textit{G5}. 
\end{itemize} 
\noindent
In the following sections, we describe the design of each \name{} component.

\subsection{Ravens (Bug Descriptors)} \label{sec:bug_context}
\name{} uses C-syntax to define \textit{Ravens}, to provide ground-truth knowledge of conditions for a bug manifestation during a fuzz input replay.\footnote{We feel the songbird, Raven, with a spiritual or godlike status in various cultures and folklore aptly describes a bug oracle role.} To accommodate the diverse nature of bugs and the complex task of identifying their manifestations and states, we leverage the introspection capabilities of the emulator to inspect the CPU’s state at specific points during execution. This process is facilitated by a Raven for each bug in a benchmark binary.  A Raven encapsulates the minimum information to uniquely identify a bug's status and is composed of two key components: i) a \textit{Reflection} section; and ii) an \textit{Introspection} section. The schema for constructing a Raven is shown in Listing~\ref{lst:Raven_Template} and we describe a collection of Ravens for a binary.

\begin{listing}[!b]
\begin{minted}
[fontsize=\footnotesize,numbersep=4pt,xleftmargin=10pt,bgcolor=lightgray,linenos,escapeinside=||]{c}
/* Reflection*/
context_struct hook_addresses[] = {
    {|\codeangle{Address}|, func_1},
    ...
};

/* Introspection*/
void func_1() {
    // Report bug Reached state
    report_reached(|\codeangle{BUG ID}|);
    // Read register of interest
    |\codeangle{Register}| = frb_reg_state(|\codeangle{Register Id}|); 
    // Read variable of interest from memory
    |\codeangle{Memory}| = frb_mem_read(|\codeangle{Address}|,|\codeangle{Size}|);
    // Evluate bug triggered state
    if (|\codeangle{Logical Expression}|) {
        //Report bug Triggered/Detected state
        report_detected_triggered(|\codeangle{BUG ID}|); 
    }
    // Bug not Reached
}
\end{minted}
\caption{Schema of a \textit{Raven}. Here \texttt{<Logical Expression>} crafted with \texttt{<Register>} and \texttt{<Memory>} values, encapsulates a bug's triggered status.
} \label{lst:Raven_Template}
\end{listing}

The reflection section defines locations along the execution path---program counter values represented as \codeangle{Address}---for bug evaluation or data collection. At each  reflection point, the associated introspection function---such as \color{blue!75!white}\texttt{func\_1()}\color{black}---is invoked to analyze relevant state. As shown in Listing~\ref{lst:Raven_Template}, these mappings are encoded in the \texttt{hook\_addresses} array using \texttt{context\_struct} elements. Each element represents a mapping between a specific program counter (PC) address and the corresponding introspection function to be triggered when execution reaches that address.

At the point of introspection, the invoked function either: immediately evaluates the bug state, or gathers introspection information---such as capturing register values or memory contents at specific program points. This is done through \texttt{frb\_reg\_state} and \texttt{frb\_mem\_read} functions to access both \codeangle{Registers} and \codeangle{Memory} contents, respectively. The collected introspection data is retained and can be referenced by the same or other Ravens later in the replay, enabling the evaluation of more complex or state-dependent bug conditions. The introspection logic invokes either \texttt{report\_reached} or \texttt{report\_detected\_triggered}; these functions output the \textit{Bug Data}---a bug’s state and identifier \codeangle{BUG ID}---. Effectively, a bug is reported as \textit{Reached} if the Raven function is called, and as \textit{Triggered} or \textit{Detected} if the associated \codeangle{Logical Expression} is satisfied. If this logical expression is satisfied by a crashing seed, the bug is recorded as \textit{Detected}. When neither the Raven is called nor the logical expression is satisfied, the bug state is inferred as \textit{Not Reached}. 

\subsubsection{Raven Construction and Validation}\label{sec:Raven-create-correct}

The workflow for crafting a Raven for a new bug builds upon existing methods for root cause analysis as follows:

\begin{itemize}[itemsep=2pt,parsep=1pt,topsep=3pt,labelindent=5pt,leftmargin=12pt]
    \item[1.] \textbf{Identification.~}New bugs manifest as ungrouped crashes (i.e., not matched/identified by an existing Raven).
    \item[2.] \textbf{Root Cause Analysis.~}Determine the underlying vulnerability, simultaneously identifying the necessary state conditions for Introspection and execution points for Reflection.
    \item[3.] \textbf{Construction.~}Encode the conditions for Introspection and execution points for Reflection into a new Raven with an assigned Bug ID using C-syntax style coding.
    \item[4.] \textbf{Verification.~}A Raven's correctness is verified by replaying crashing seeds. A correct Raven must: i)~reliably detect the target bug using the intended input (a Raven covers all related crashes); ii)~not falsely report distinct bugs (i.e. seeds associated with previously known bugs are not captured by the new Raven); and iii)~ensure no crashes remain unlabeled (no ungrouped crashes remain).
\end{itemize}

Then, we can consider a set of Ravens for a target to be complete if there are no unidentified crashing seeds. Consequently, \name{} aims to achieve no unidentified crashes during the analysis of a target, unless it is a bug for which a Raven remains to be defined.

Describing a bug as a Raven demands a thorough understanding of its root cause through static and dynamic analysis. While root cause analysis is time-consuming (typically ranging from 30 minutes to 4 hours in our experience), it is a standard prerequisite for responsible bug disclosures. Crucially, the information required for a Raven---the Reflection points and Introspection conditions---is naturally uncovered during this debugging phase. As a result, the subsequent construction of a Raven is generally straightforward, taking approximately 30 minutes.

While automating the root cause analysis process is orthogonal to our work, we have spent considerable effort to curate a comprehensive collection of pre-defined Ravens covering a wide range of binaries grouped into three benchmark sets (\namebench{}, \namebenchdma{} and \namebenchX{}), as discussed later in Section~\ref{sec:BechnmarkTargets}. Importantly, after root cause analysis of a bug, we have made the process of incorporating a bug with a Raven, simple---we demonstrate the ease with which Ravens can be constructed using three case examples of common bug types:
\begin{itemize}[itemsep=2pt,parsep=1pt,topsep=3pt,labelindent=5pt,leftmargin=12pt]
    \item Type Confusion
    \item Stack Buffer Overflow (deferred to our code repository~\cite{FirmReBuggerGit})
    \item Dangling pointer (deferred to our code repository~\cite{FirmReBuggerGit})
\end{itemize}

\begin{listing}[!b]
\begin{minted}
[fontsize=\footnotesize,numbersep=4pt,xleftmargin=10pt,bgcolor=lightgray,linenos]{c}
context_struct hook_addresses[] = {
    {0x08005e28, BUG_MF04},
    ...
}

void BUG_MF04() {
    report_reached("MF04");
    //canbus fail to verify device type
    uint32_t read_addr = frb_reg_state[0] + 0x4;
    if (frb_mem_read(read_addr,4) != 0x0800f7e4){
        report_detected_triggered("MF04");
    }
}
\end{minted}
\caption{Type Confusion example. A Raven crafted for the bug "MF04" in \bin{Zephyr SocketCan} binary from \uemu{}~\cite{uEmu}.}\label{lst:context_mem_reads}
\end{listing}

\noindent\textbf{Type Confusion.~} This arises when a program erroneously interprets an incorrect type for a region of memory. As a consequence, the program may access fields, invoke functions, or perform operations that are invalid for the underlying data, leading to undefined behavior. Introspection of object types and pointer usage can help identify type confusion bugs.

\noindent\textbf{\textit{Raven Example.~}} An illustrative example of a Raven is shown in Listing \ref{lst:context_mem_reads}, based on the bug ID \texttt{MF01} reported by \multifuzz{}~\cite{chessermultifuzz} in the \bin{Zephyr SocketCAN} binary \cite{uEmu}. In Zephyr's device model, each device is represented by a struct containing a pointer named \texttt{driver\_api}. This pointer references a table of function pointers that define the operations supported by the device's driver, such as configuration or data transmission. The CAN bus subcommands allow users to specify a target device for command execution. At runtime, the target device is resolved using the \texttt{z\_impl\_device\_get\_binding} function, which returns a pointer to a generic device struct. However, no type verification is performed to ensure that the selected device implements the CAN bus API. As a result, if a non-CAN device is specified (such as a GPIO device), the subcommand will erroneously perform CAN bus operations on an incompatible device struct.

To capture this bug in Raven, the key condition to verify is whether \texttt{dev->driver\_api} at the point it is used, points to a valid CAN bus API function table. In this context, as shown in Listing~\ref{lst:context_mem_reads}, whether it matches the address \texttt{0x0800f7e4}, which is the base address of the CAN driver’s function table. This check should be performed at the point in the CAN bus subcommand function where \texttt{dev->driver\_api} is about to be invoked (\texttt{0x08005e28}). Since \texttt{dev->driver\_api} is not conveniently held in a register prior to access, its value must be read directly from memory. The introspection function \texttt{frb\_mem\_read} is used to read a specified memory address with the appropriate byte width. In the context of Listing~\ref{lst:context_mem_reads}, reading from \texttt{R0 + 0x4} yields the value of \texttt{dev->driver\_api}, which can then be compared to the expected CAN API function table address to determine whether the bug is triggered.

We systematically develop \numoracles{} accurate Ravens for benchmarking using the method described here by analyzing the bugs in target binaries we discuss in Section~\ref{sec:BechnmarkTargets}.

\subsection{Bug Interpreter}
The \textit{Bug Interpreter} orchestrates the evaluation of each bug during the replay of fuzz inputs---including crashing inputs. By interpreting Ravens with dynamic CPU state observations for every input, the \textit{Bug Interpreter} identifies and records the corresponding bug state and Bug ID. This design decouples the Raven logic from the emulator itself, enabling updates or extensions to Ravens without modifying either the binary under test or the emulator’s internal logic.

At the start of each replay session, the \textit{Bug Interpreter} is initialized by dynamically loading Ravens and configuring the emulator with hooks at the reflection points specified by each Raven. During replay, whenever the program counter reaches one of these hooked locations, the \textit{Bug Interpreter} activates the corresponding hook and transfers control to the introspection section. The hook invokes the relevant Raven's logic, potentially leveraging previously gathered introspection data or the current CPU state to assess the bug's status. The resulting \textit{Bug Data}---bug state and corresponding ID---are recorded and forwarded to the \textit{Analysis Bench} to generate bug-based metrics. 

Importantly, the Ravens and the Bug Interpreter are agnostic to a target emulator. We provide an API to connect the Bug Interpreter to a given emulator and implement the \textit{Bug Interpreter} using TinyCC \cite{tinycc}. Notably, while many fuzzing approaches exist, the emulators employed are limited to QEMU~\cite{bellard2005qemu}, Unicorn~\cite{quynh2015unicorn} (ported from QEMU) and Icicle~\cite{chesser2023icicle}. We provide implementations of our Bug Interpreter API to connect to each of these emulators. Collectively, the attributes supported simplify the adoption of our framework for researchers and developers while Ravens are \emph{portable} across fuzzers.

\subsection{Analysis Bench}
The \textit{Analysis Bench} assimilates the Bug Data generated by the \textit{Bug Interpreter}, shown in Figure~\ref{figure:overview}, to generate evaluation metrics.  As \name{} operates purely as a post-processing step, it can incorporate fuzzing logs from the campaign to extract crash timing information necessary for deriving time-to-bug statistics. The \textit{Bug Analyzer} determines the state of each known bug (not reached, reached, triggered or detected) for every fuzz input and computes the key evaluation metrics such as the \textit{time to reach a bug}, \textit{time to trigger a bug}, and \textit{the number of unique bugs triggered} used in bug-based fuzzing benchmarks~\cite{zhang2022fixreverter, hazimeh2020magma}. Since the program’s state cannot be guaranteed after a bug is triggered, only the first triggered bug per input is reported. The rationale for this choice is because a firmware may enter a compromised state after triggering a bug, making subsequent bug reports unreliable (i.e., later reports may depend on the invalid state produced by the first bug). To facilitate further investigation, the bug metrics generated by \textit{Bug Analyzer} flags any crashes that triggered multiple bugs. The analysis generates metrics in a standard, machine-readable format, so the \textit{Charting Tool}---which parses data to generate visualizations and summary statistics---can function as an independent, reusable component for fuzzers or future fuzzing frameworks.

{\begin{mdframed}[backgroundcolor=blue!10,rightline=false,leftline=false,topline=false,bottomline=false,roundcorner=2mm] 
    Importantly, once Ravens are added, Ravens and the Analysis Bench are portable across all fuzzers.
\end{mdframed}} 

\section{FirmBench: Benchmark Targets}\label{sec:BechnmarkTargets}
This section examines the common roadblocks faced by current SoTA monolithic firmware fuzzers. Our findings guide the development of our benchmark target sets.

\subsection{Common Roadblocks} 
\label{Common Roadblocks in Fuzzing Firmware}
Roadblocks are challenges that hinder a fuzzer's progress, reduce throughput, restrict access to key sections of code, or prevent accurate injection of data from the fuzzer into the firmware. Monolithic firmware is riddled with various roadblocks, some commonly encountered in other software such as \textit{magic values}~\cite{aschermann2019redqueen} but difficult solve in the firmware domain~\cite{farrelly2023splits} or those unique to firmware such as interactions with DMA~\cite{dice2021,zhou2022semu,scharnowskigdma}.

We synthesize  a set of common monolithic firmware fuzzing roadblocks by analyzing the problems reported and encountered in the nine state of the art fuzzers~\cite{dice2021,zhou2022semu,scharnowskigdma,feng2020p2im,scharnowski2022fuzzware,farrelly2023splits,uEmu, scharnowski2023hoedur,chessermultifuzz}. 
This involved: 
\begin{itemize}[itemsep=2pt,parsep=1pt,topsep=3pt,labelindent=5pt,leftmargin=12pt]
    \item[1)] Reviewing each of the papers to identify aspects considered out-of-scope and/or unaddressed challenges and limitations.
    \item[2)]\textit{Crucially}, the practical workarounds found for fuzzing binary targets in associated code repositories, such as source patches, feature selection (e.g., build-flags), and code assumptions.
\end{itemize}

\noindent 
We discuss the common roadblocks identified, below.
\begin{listing}[h]
\begin{minted}
[fontsize=\footnotesize,numbersep=4pt,xleftmargin=10pt,bgcolor=lightgray,linenos]{cpp}
...
    iVar1 = strncmp(argv[1],"settime",7);
    if ((iVar1 != 0) || (argc != 4)) {
      iprintf("unknown command...",argv[1]);
      _rtc_usage();
      return 1;
    }
// Critical Code Block
\end{minted}
\caption{String Comparison in the \bin{Console} Binary \cite{feng2020p2im}}\label{lst:stringcmp}
\end{listing}

\vspace{1mm}
\noindent \textbf{Magic Values.}
Magic values are special constants such as specific strings, numbers or byte sequences that programs use to identify formats, states or to trigger particular behaviors. For fuzzers, these values act as roadblocks, because they enforce strict input checks that must be satisfied before deeper program logic can be reached \cite{aschermann2019redqueen}. 

This problem is even more pronounced in firmware fuzzing, where inputs are often processed one byte at a time from various peripheral registers through interactions such as interrupts, rather than as contiguous blocks (such as files). This fragmentation makes it more difficult for firmware fuzzers to identify and solve the magic values to make progress~\cite{farrelly2023splits}. A example from real-world firmware is shown in Listing~\ref{lst:stringcmp}, where a string comparison guards critical code. In prior work \cite{feng2020p2im,scharnowski2022fuzzware,zhou2022semu,scharnowski2023hoedur}, source-level patches were applied to bypass such checks in some targets to remove checksums or simplify complex comparisons, confirming that they remain a significant roadblock for state-of-the-art fuzzers.

\vspace{1mm}
\noindent \textbf{Complex Peripherals.}
Fuzzing complex peripherals such as USB or SPI presents unique challenges because both require highly structured, protocol-compliant inputs. Unlike simple data register (DR) accesses, these interfaces demand sequences of interactions that adhere to specific packet formats, command-response patterns, and internal state transitions. Because fuzzers typically generate context-free inputs, the resulting sequences are often infeasible on real hardware. As a result, malformed or incomplete transactions are usually discarded before exercising deeper functionality. In some cases, invalid sequences may trigger crashes that would never occur on real hardware, leading to false positives and reduced fuzzer throughput.

\begin{figure}[t!]
    \centering
    \includegraphics[width=\linewidth]{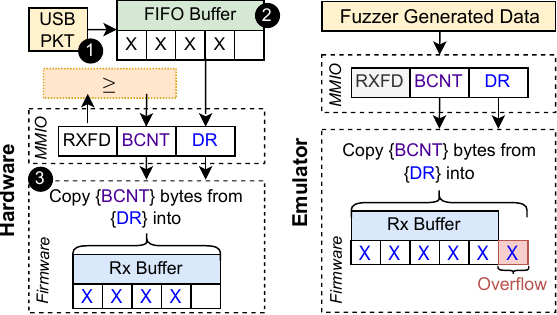}
    \caption{Example USB peripheral interaction in hardware vs. emulation. buffer. 
     On hardware, BCNT is always limited by RXFD. In emulation, this check is not enforced, allowing a buffer overflow not possible on the target hardware during the copy.}
    \label{figure:usb}
\end{figure}

An example interaction with a complex peripheral, USB, where the hardware behavior difficult to model in an emulated environment is shown in Figure~\ref{figure:usb}. The firmware configures the RXFD register to set the size of the hardware FIFO buffer and allocates a matching RX buffer in RAM. When a USB packet arrives \blackcircled{1}, it is placed in the FIFO buffer. An interrupt signals \blackcircled{2} the firmware to  read the BCNT (byte count) value from the FIFO Buffer, via the DR register, and copies BCNT-bytes into the RX buffer \blackcircled{3}. The hardware enforces BCNT never exceeds RXFD---and thus the size of the RX buffer---allowing the firmware to assume the value is in bounds. In the emulator, the fuzzer can provide arbitrary or out-of-bounds values for BCNT, ignoring the constraints imposed by the hardware ($\mathrm{RXFD} \geq \mathrm{BCNT}$). As a result, the firmware may copy more data than the RX buffer can hold, leading to a buffer overflow that is not possible on real hardware.

Prior work \cite{feng2020p2im,zhou2022semu,scharnowski2022fuzzware, scharnowski2023hoedur} applied source-level patches to simplify or bypass these interactions, such as replacing SPI communication with simplified direct MMIO reads, simplifying BLE and RF chip interactions, or disabling specific interrupts.

\begin{figure}[!b]
    \centering
    \includegraphics[width=\linewidth]{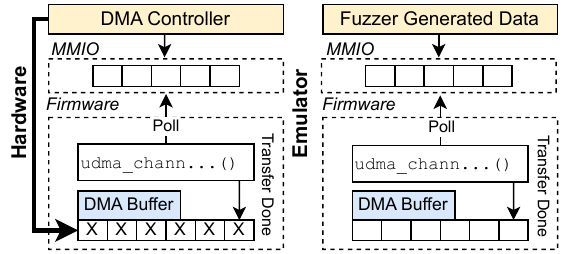}
    \caption{Comparison of DMA in hardware and a fuzzer's emulated environment. The emulator's buffer remains empty.}
    \label{figure:dma2}
\label{lst:dma}
\end{figure}

\noindent \textbf{DMA.}
Fuzzing firmware that uses DMA (Direct Memory Access) is challenging because transfers occur between memory and peripherals without CPU involvement. This makes it difficult for fuzzers to determine when DMA transfers should be triggered and where the associated data needs to be injected. 

Figure~\ref{figure:dma2} illustrates these challenges. Hardware populates DMA buffers while firmware polls MMIO registers using the \texttt{udma\_chann...()} function to detect completion. Fuzzers may set MMIO values or trigger interrupts to indicate transfer completion, but without populating the DMA buffer, preventing exploration of DMA-dependent code paths.

DMA is frequently listed as a limitation in prior work, and while some recent studies~\cite{scharnowskigdma,zhou2022semu} have proposed techniques to handle DMA interactions, the inherent complexity still poses significant challenges for effective fuzzing.

\begin{listing}[h]
\begin{minted}
[fontsize=\footnotesize,numbersep=4pt,xleftmargin=10pt,bgcolor=lightgray,linenos]{cpp}
__weak void HAL_Delay(__IO uint32_t Delay) {
  uint32_t tickstart = 0;
  tickstart = HAL_GetTick();
  while((HAL_GetTick() - tickstart) < Delay) {
  }
}
\end{minted}
\caption{Example of a delay in the \bin{Drone} target \cite{feng2020p2im}.}\label{lst:delay}
\end{listing}

\vspace{1mm}
\noindent\textbf{Execution Delays and Input Bloating.}
A major challenge in firmware fuzzing is handling encounters with execution delays and demand for longer inputs. Often, firmware intentionally include various types of delays to synchronize with hardware or wait for events. However, in the context of re-hosted emulation-based fuzzing, these delays reduce fuzzing throughput potentially becoming a roadblock.

Another related issue is the growth in fuzz inputs. Input sizes can balloon due to various problems in the fuzzing process, such as inaccurate MMIO emulation and improper handling of interrupts. Larger inputs reduce the fuzzing efficiency as these slow execution and make it more difficult to explore the input space effectively.

A real-world example of the input bloating problem is observed in Listing~\ref{lst:delay}. Each call to the \texttt{HAL\_Delay} function requires a large number of interrupts to be triggered to advance the tick counter. The impact is two-fold: i)~it reduces fuzzing throughput; and ii)~if there are MMIO accesses within the triggered interrupt handlers, it inflates the input size needed to make progress.

To mitigate the performance and input bloat effect of execution delays, prior works \cite{feng2020p2im, uEmu, zhou2022semu, farrelly2023splits, scharnowski2022fuzzware} have adopted various strategies. These include applying source-level patches to shorten or eliminate delay loops, as well as implementing emulation heuristics to detect and fast-forward through delay polling loops.

\begin{figure*}[b!]
    \centering
    \includegraphics[width=\textwidth]{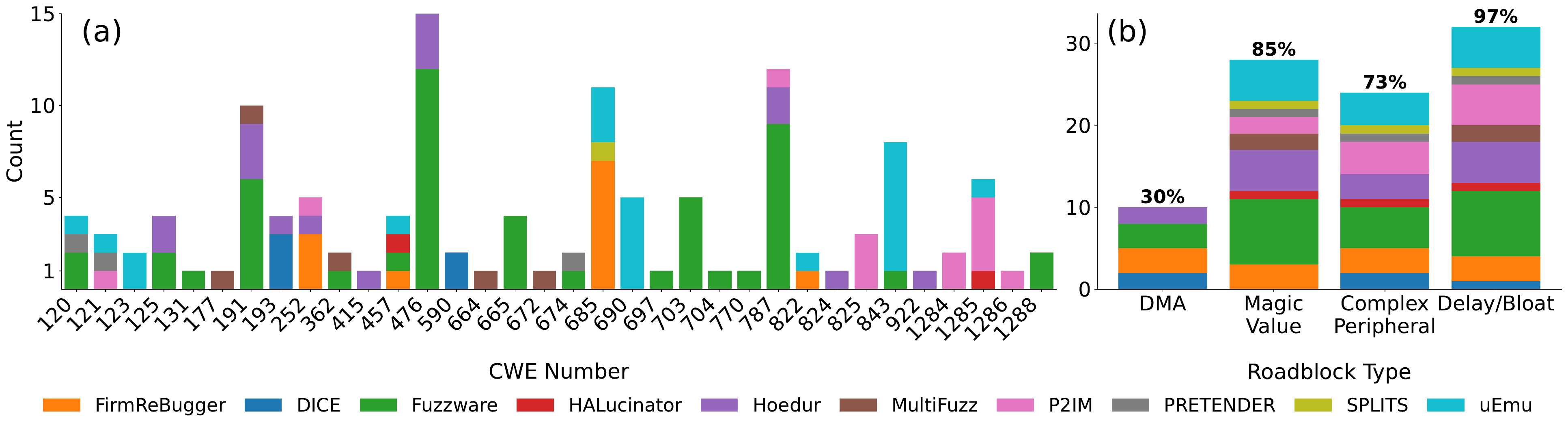}
    \caption{(a) Distribution of CWEs across unique bugs, excluding FPs and (b) roadblocks across the sample of \baseNumBin{} binaries.}
    \label{figure:combined_diversity}
\end{figure*}

\subsection{Benchmark Target Sets}
Considering the fuzzing challenges we discussed, the need to support incremental progress in the field, and and future advances, we curate three benchmarks.
To construct a suitable benchmark, we create a set of binaries that not only align with the design goals \textbf{\textit{G3}} and \textbf{\textit{G4}} described in Section \ref{sec:OurWork}, but also satisfy the following criteria:
\begin{itemize}[itemsep=2pt,parsep=1pt,topsep=3pt,labelindent=5pt,leftmargin=12pt]
    \item \textbf{Verifiable.~}The binary must contain a confirmed bug, with evidence that the bug is reachable.
    \item \textbf{Non-Trivial.~}The bug should be sufficiently complex (>1 hour to be triggered by any fuzzer).
    \item \textbf{Monolithic.~}The binaries within the \namebench{} set must be monolithic.  
\end{itemize}
\noindent 
The benchmark binary selection draws from a comprehensive review of prior state-of-the-art monolithic firmware fuzzing efforts, further augmented by  additional large binaries containing newly identified bugs. We examined all real-world binaries spanning nine published studies to develop the target sets. The collection incorporates an analysis of 50
    binaries introduced in SoTA fuzzers, \ptwoim{}~\cite{feng2020p2im}, \textsc{PRETENDER}~\cite{gustafson2019Pretender}, \uemu{}~\cite{uEmu}, \textsc{HALucinator}~\cite{clements2020halucinator}, \fuzzware{} \cite{scharnowski2022fuzzware}, \dice{}~\cite{dice2021}, \splits{}~\cite{farrelly2023splits}, \hoedur{}~\cite{scharnowski2023hoedur}, and \multifuzz{}~\cite{chessermultifuzz}. We selected a sample of 30 binaries from the 50 we analyzed in prior works following the criteria above.
Further, we expand this set with three new large binary targets containing 15 newly introduced bugs and the discussed roadblocks to curate a total of \baseNumBin{} base binaries. The \baseNumBin{} binaries selected contain \numuniquebugs{} unique software bugs we employed to construct the three benchmark target sets detailed below. Importantly, by carefully and systematically analyzing each bug, we developed accurate Ravens for every bug in each benchmark set using the methods in Section~\ref{sec:firmrebugger}.

\begin{itemize}[itemsep=2pt,parsep=1pt,topsep=3pt,labelindent=5pt,leftmargin=12pt]
    \item \namebench:~To establish a standardized baseline for fuzzing comparisons, without the impediments from the roadblocks, we include 28 binaries selected from prior work. 
    Additionally, we incorporate the three large binaries introduced in this work; consistent with prior binaries, we apply source-level patching to remove challenging roadblocks. The curated dataset comprises of 166 bugs, each with a corresponding Raven, spread across 31 binary targets. 
    \item \namebenchdma:~Managing DMA injected data remains a key challenge, this set of binaries aim to support the development of fuzzers addressing the problem. The set is built by filtering the \baseNumBin{} selected binaries to include only those containing DMA interactions and reverting patches that replaced DMA interactions with equivalent MMIO operations in those binaries included from prior work. The curated set comprises 31 bugs, each with a corresponding Raven, spread across 8 binary targets.
    \item \namebenchX. In our benchmarking efforts, we recognize that the use of fuzzing-specific binaries---where certain challenges are patched or removed---can unfairly disadvantage fuzzers designed to tackle these very problems. To address this, we introduce a challenging benchmark set, \namebenchX{}, in which binaries are left untouched with respect to the roadblocks we discussed. We make patches/modifications only when absolutely necessary to enable the binary to execute in an emulator. These patches address emulator capability limitations (unrelated to the roadblocks), such as substituting any hardware-assisted implementations of functions like \texttt{memcpy} with their software equivalents. This policy of minimal intervention preserves the inherent complexity of the original binaries. The binaries curated offer a challenging, realistic and rigorous benchmark for future advances in fuzzers. The curated set comprises 109 bugs, each with a corresponding Raven, spread across 22 binary targets.
\end{itemize}

Notably, we retain false positive bugs. A false positive occurs when a fuzzer reports a crash that arises from an emulated state that cannot occur on real hardware. Although false positive bugs are undesirable, they can demonstrate a fuzzer's ability to reach deep execution paths and uncover latent vulnerabilities~\cite{scharnowski2022fuzzware}. We clearly identify false positives, discussing them in detail in Appendix~\ref{appendix:false-positives}.

\begin{figure*}[b!]
    \centering
    \includegraphics[width=1\textwidth]{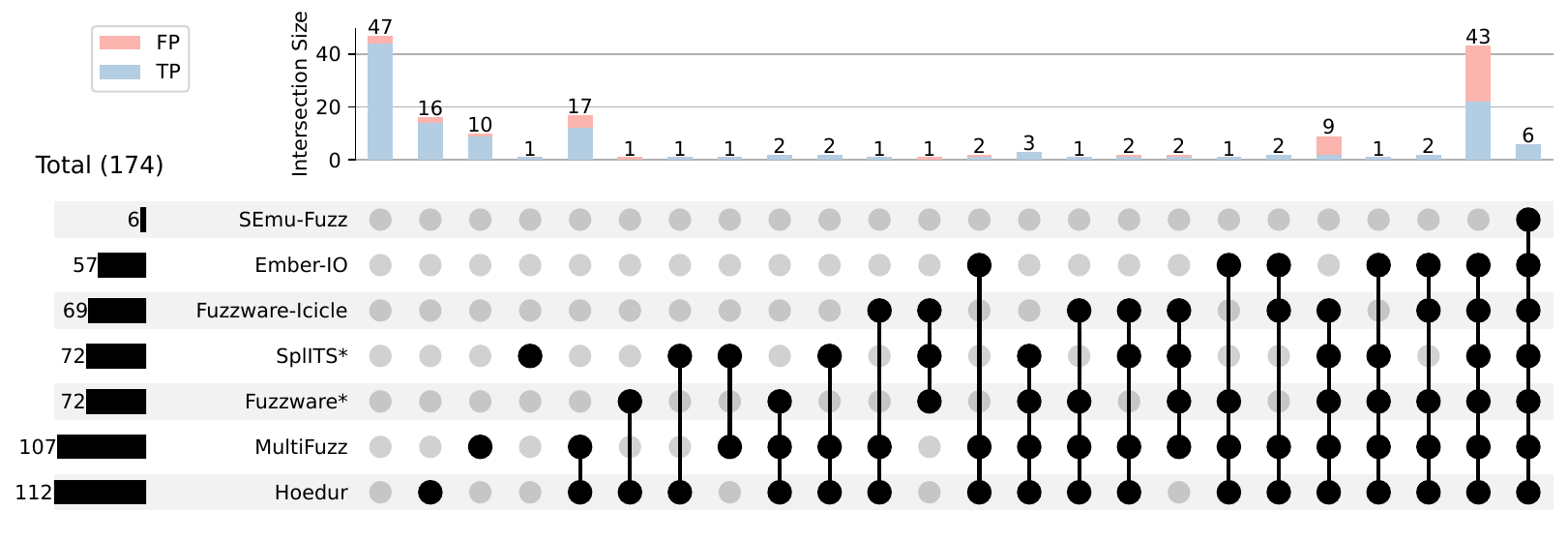}
    \caption{Distribution of true and false positive bugs across fuzzers over 10 trials (24h) on the \namebench{} benchmark set. Note: fuzzers marked with an asterisk (*) were evaluated on a subset of binaries--we detail our rationale in Appendix~\ref{appendix:evaluation-setup}.}
    \label{figure:firmbenchupset}
\end{figure*}

\subsubsection{Analysis of the Benchmark Binary Targets}\label{sec:diversity}

\namebench{} encompasses \numuniquebugs{} unique bugs and \numoracles{} Ravens across \uniqueCWECount{} distinct CWEs, as summarized in Figure~\ref{figure:combined_diversity}(a). To analyze the benchmark, for each bug, we assign a single CWE following the guidelines provided by MITRE~\cite{mitreguidelines}. In cases where bugs have associated CVEs, we adopt the CWE specified in the corresponding CVE record. The selected \baseNumBin{} binaries along with the roadblocks reflected in these binaries, and any associate bug description CWEs are summarized in Table~\ref{table:summary_binaries}. 
We detail the modifications made to the selected binaries from prior work to construct the \numbin{} binaries (\namebench{}:~31, \namebenchdma{}:~8, \namebenchX{}:~22) used in the benchmark sets in our code repository~\cite{FirmReBuggerGit}.

As illustrated in Figure~\ref{figure:combined_diversity}(b), the \numuniquebugs{} unique bugs in \baseNumBin{} firmware targets exhibit diverse bug categories (34 different CWEs) and with significant representations of roadblocks: 28 (85\%) implement magic-value checks, 24 (73\%) interact with complex peripherals, 10 (30\%) leverage DMA, and 32 (97\%) contain delay or bloat code. Collectively, these targets cover 14 distinct microcontroller and 10 unique system libraries 
as detailed in Table~\ref{table:summary_binaries}. Notably, the analysis reflects the current state of the benchmark but \name{} supports the community to expand the benchmark sets as new bugs are discovered.

\section{Experimental Evaluations}\label{sec:experiments}
We evaluated the performance of state-of-the-art fuzzers to establish the versatility of our benchmark suite and understand the effectiveness of techniques explored in recent studies. 

\vspace{1mm}
\noindent
\textbf{Fuzzers.~}We selected nine state-of-the-art firmware fuzzers for evaluation: \fuzzware{} \cite{scharnowski2022fuzzware}, \ember{} \cite{farrelly2023ember},  \fuzzwareicicle{} \cite{chesser2023icicle}, \semu{} \cite{zhou2022semu}, \dice~\cite{dice2021}, \gdma{}~\cite{scharnowskigdma}, \splits{} \cite{farrelly2023splits}, \hoedur{} \cite{scharnowski2023hoedur}, and \multifuzz{} \cite{chessermultifuzz}. Notably, not all targets are executed on \semu{} since it requires manual, platform-specific setup for new targets. Additionally, due to unimplemented instructions within the version of Unicorn~\cite{quynh2015unicorn} used by \fuzzware{} and \splits{}, we exclude the \bin{Oresat-Control} binary from the evaluation of these fuzzers. Then, for \namebenchdma{}, we employ the recent fully-automated fuzzers~\cite{scharnowskigdma,dice2021} focusing on extending the existing firmware fuzzers to support DMA using information gathered at runtime\footnote{Although a concurrent study to \gdma{} investigating fully automated methods of data injection to DMA, \dyma~\cite{Dymafuzz}, was recently published, at the time of benchmarking, the method was unavailable. We defer its integration to the benchmark to our code repository.}. We defer extended details of the evaluation rationale and settings to Appendix~\ref{appendix:benchmark-settings}. 

\begin{figure*}[t!]
    \centering
    \includegraphics[width=\textwidth]{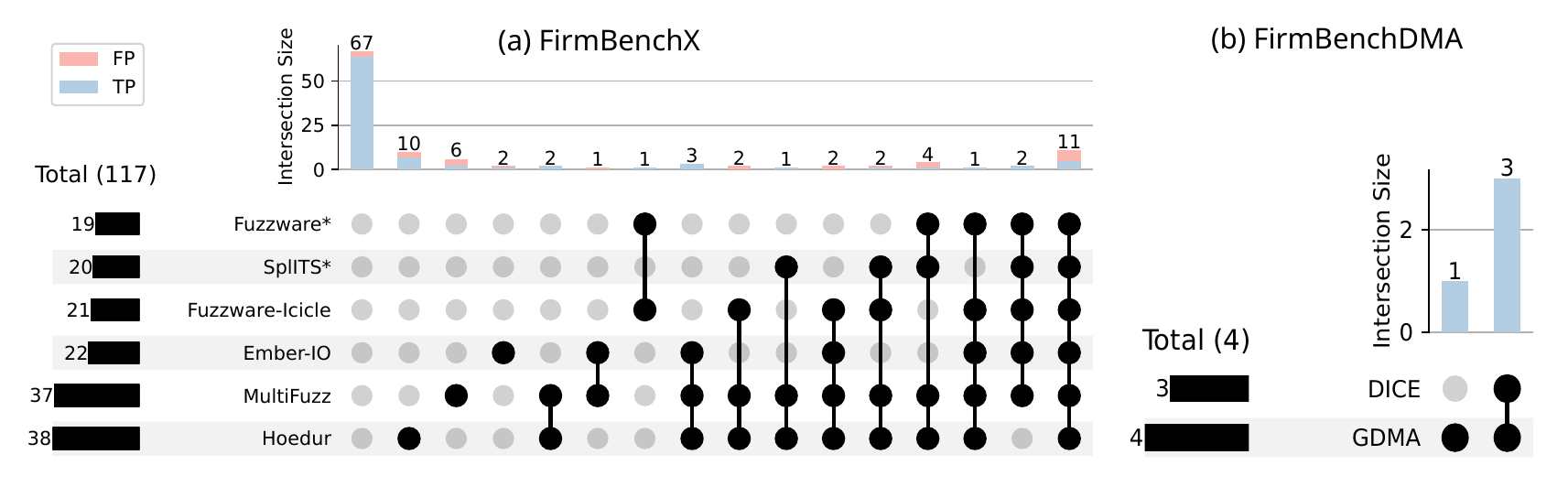}
    \caption{Distribution of true and false positive bugs across fuzzers over 10 trials (24h) on the (a)~\namebenchX{} and (b)~\namebenchdma{} benchmarks. Fuzzers marked with an asterisk (*) were evaluated on a subset of binaries---see Appendix~\ref{appendix:evaluation-setup}.}
    \label{figure:upsetX}
\end{figure*}

\begin{figure}[b!]
    \centering
    \includegraphics[width=\columnwidth]{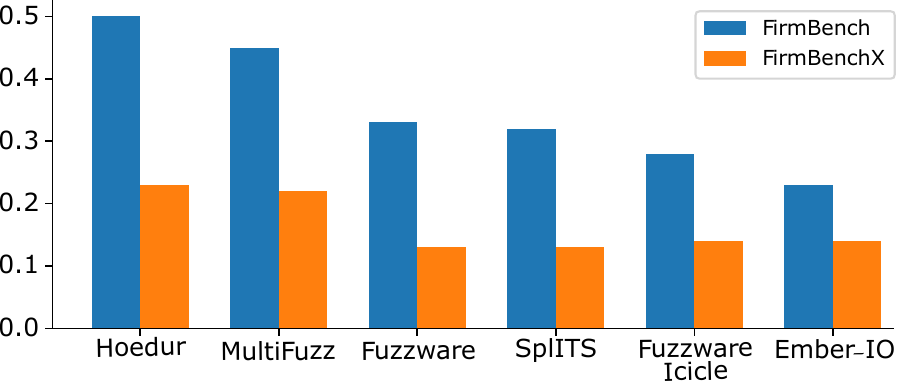}
    \caption{Consistency score comparison. Higher scores generally indicates greater bug-finding reliability. But it should be interpreted alongside the number of bugs triggered as consistency alone does not capture the breadth of bug coverage.}
    \label{figure:consistency}
\end{figure}

\vspace{1mm}
\noindent
\textbf{Benchmark Metrics.~} We report: i)~\emph{bug count}---the number of unique bugs triggered at least once across 10 independent trials; ii)~\emph{time-to-bug} and survivability functions following the method in Magma~\cite{hazimeh2020magma}; and iii)~we evaluate each fuzzer’s \emph{consistency}
to measure the reliability with which bugs are discovered across repeated trials 
where
consistency is defined as:
\[
\text{Consistency}(f) = \frac{1}{|B|} \sum_{b \in B} \left( \frac{c_{f,b}}{T} \right),
\]

\noindent
where, $f$ is a particular fuzzer, $B$ is the set of all bugs, $|B|$ is the total number of bugs, $c_{f,b}$ is the number of trials in which fuzzer $f$ triggered bug $b$, and $T$ is the total number of trials.

\noindent
\textbf{Evaluation Regime.~}
We execute each fuzzer for 24-hours and use 10 trials on each binary target. This is a significant effort, totaling 10 CPU-years of computation time. 

\section{Results and Discussions}

In this section, we discuss the versatility and utility of \name{} as a benchmark, and investigate the effectiveness of various existing fuzzers based on our results.

Figure~\ref{figure:firmbenchupset} and \ref{figure:upsetX} summarize the bug finding ability of fuzzers in upset plots; the vertical bars represent the number of triggered bugs in each intersection, the horizontal bars show the total number of bugs triggered per fuzzer, and the connected dots denote the fuzzers involved in each intersection. Figure~\ref{figure:consistency} illustrates consistency, Figure~\ref{figure:survivial} presents a selection of survivability plots with Table~\ref{appendix:firmbenchdma01} showing the generated time-to-bug metrics for \namebenchdma{} (full results are available on our GitHub repo~\cite{FirmReBuggerGit}).

\subsection{Observations}

\noindent
\textbf{Overall Analysis.~}Across our three benchmark sets, comprising a total of 295 bugs, the evaluated fuzzers triggered 181 bugs overall: 127 of 174 in \namebench{}, 50 of 117 in \namebenchX{}, and 4 of 4 in \namebenchdma{}. As shown in Figure~\ref{figure:firmbenchupset} and Figure~\ref{figure:upsetX}~(a), \multifuzz{} and \hoedur{} are the best performing fuzzers, triggering 144 and 150 bugs, respectively. In contrast, single-streamed fuzzers performed notably worse, with similar results to each other---except for \semu{}. \splits{} performed marginally better than the others by uncovering bugs guarded by solving magic value comparisons.

\begin{figure*}[t]
    \centering
    \includegraphics[width=\textwidth]{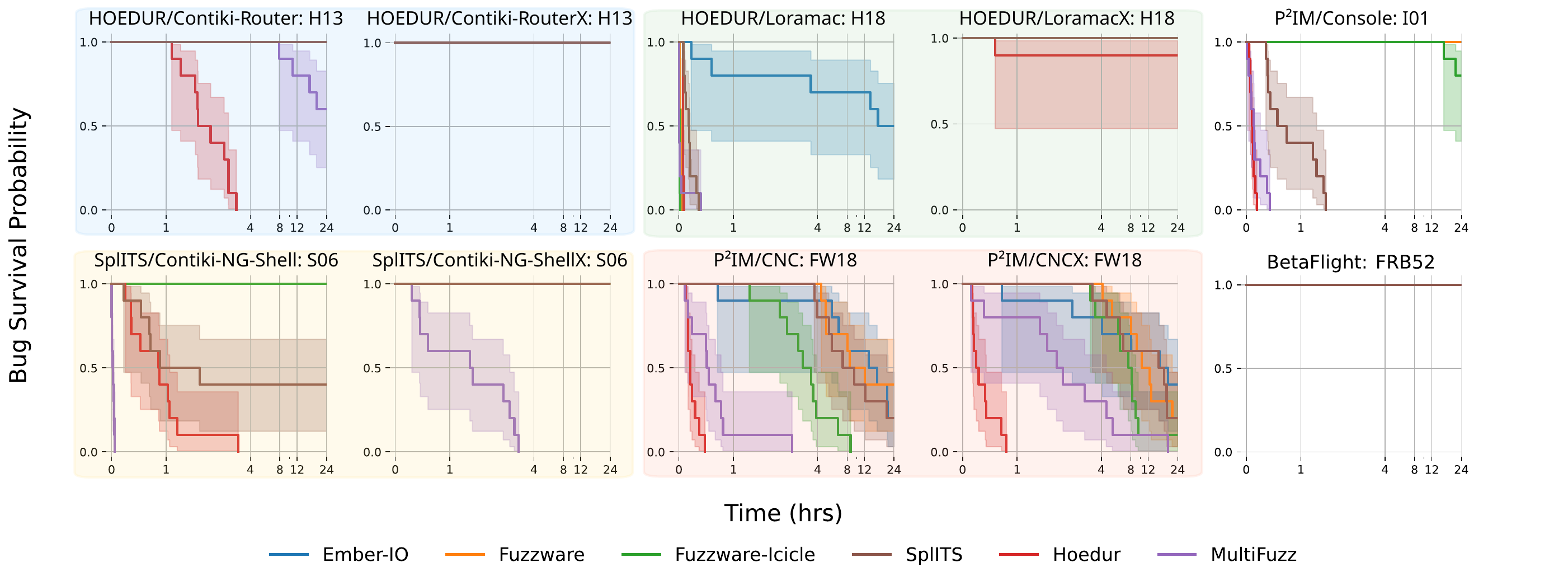}
    \caption{Survival functions for selected bugs. The shaded areas indicate 95\% confidence intervals. Colors group pairs of related binaries (e.g., original and ‘X’ variant) for visual comparison (statistics and plots generated by our Analysis Bench---see Fig.~\ref{figure:overview}).}
    \label{figure:survivial}
\end{figure*}

Figure~\ref{figure:consistency} summarizes the consistency of each fuzzer at triggering bugs. Comparing the two leading fuzzers, overall, \hoedur{} demonstrates higher consistency than \multifuzz{}. We hypothesize that \hoedur{}'s improved consistency may be due to its use of probability matching (Thompson Sampling~\cite{chapelle2011empirical}) to adaptively select data streams for mutation based on their observed success rates in increasing coverage~\cite{scharnowski2023hoedur}. This approach aims to avoid being misguided by frequent but uninformative register accesses.

Our evaluation on \namebenchX{} in Figure~\ref{figure:upsetX}~(a) shows a significant drop in bugs triggered---from 73\% in \namebench{} to 43\% in \namebenchX{}. This stark decrease highlight a research gap for future developments to tackle the challenges and roadblocks we discussed in Section~\ref{Common Roadblocks in Fuzzing Firmware}.

\begin{mdframed}[backgroundcolor=blue!10,rightline=false,leftline=false,topline=false,bottomline=false,roundcorner=2mm] 
    \textbf{Key Takeaway}: 73\% of the bugs within \namebench{} were triggered by at least one fuzzer within 24 hours, while only 43\% of the \namebenchX{} bugs were triggered. As expected, recent fuzzers employing multi-stream input representations~\cite{scharnowski2023hoedur,chessermultifuzz} yielded the best performance while stream-specific mutations may improve consistency.
\end{mdframed}

\noindent
\textbf{Magic Values.~}As shown in Figure~\ref{figure:survivial}, for the \bin{Console} binary and bug I01, only \hoedur, \multifuzz, \splits, and \fuzzwareicicle{} are able to trigger the bug, with \fuzzwareicicle{} doing so only two times, while the others achieve full coverage. \fuzzwareicicle{} uses Compare Coverage (CompareCov) feedback to guide mutations near comparison instructions, which enables it to occasionally solve guarded magic values checks, but with less consistency than other approaches. In contrast, \hoedur{} combines a dictionary-based mutation strategy but with its multi-stream approach, it is able to efficiently generate and match magic value sequences more often. Notably, both \splits{} and \multifuzz{} utilize input-to-state feedback mechanisms tailored for firmware, enabling them to closely correlate input mutations with program state changes and systematically overcome magic value conditions. This is a likely source of bugs uniquely identified by each of these fuzzers (Figure~\ref{figure:firmbenchupset}).

\begin{mdframed}[backgroundcolor=blue!10,rightline=false,leftline=false,topline=false,bottomline=false,roundcorner=2mm] 
    \textbf{Key Takeaway}: Magic values are used in firmware and can guard bugs. Input-to-state mechanisms~\cite{farrelly2023splits, chessermultifuzz} or dictionaries (combined with multi-stream inputs)~\cite{scharnowski2023hoedur} prove effective at consistently overcoming these roadblocks.
\end{mdframed}

\noindent
\textbf{Complex Peripherals.~}As shown in Figure~\ref{figure:survivial}, the \bin{Loramac} and \bin{LoramacX} binaries with bug H18 differ significantly in their SPI handling. In \bin{Loramac}, the binary is patched with a direct SPI read in the \texttt{transceive} function, eliminating explicit interaction with the peripheral, whereas in \bin{LoramacX} this patch is reverted bring the interaction back. This makes fuzzing substantially more challenging for two reasons: it introduces a dependency on realistic peripheral behavior---something generic fuzzers often struggle to emulate---and it requires the fuzzer to supply specific, meaningful input via the SPI interface to trigger the bug. This not only expands the input space but also increases the complexity of exercising the bug condition. As a result, the challenge binary (\bin{Loramac}) is significantly harder for fuzzers: as evidenced by the survivability graph in Figure~\ref{figure:survivial}, only \hoedur{} was able to trigger the bug—and only once out of ten trials---whereas previously, the majority of fuzzers could do so consistently. 

\begin{mdframed}[backgroundcolor=blue!10,rightline=false,leftline=false,topline=false,bottomline=false,roundcorner=2mm] 
    \textbf{Key Takeaway}: Complex peripheral interactions prevent many fuzzers from reaching bugs in a timely manner; impacting even the most recent multi-stream methods.
\end{mdframed}

\begin{table}[h]
\centering
\resizebox{\columnwidth}{!}{%
\begin{tabular}{l|l|ccc|ccc}
\hline
Binary & Bug ID & \multicolumn{3}{c|}{GDMA~\cite{scharnowskigdma}} & \multicolumn{3}{c}{\dice~\cite{dice2021}} \\
 & & \multicolumn{3}{c|}{{Med. R \textbar{} Med. T \textbar{} Hit}} & \multicolumn{3}{c}{{Med. R \textbar{} Med. T \textbar{} Hit}} \\ \hline
MIDI-Synthesizer & DI4 & 00:13 & 00:13 & \cellcolor[HTML]{63BE7B}100\% & \textbf{00:08} & \textbf{00:08} & \cellcolor[HTML]{63BE7B}100\% \\
 & DI5 & \textbf{00:15} & \textbf{00:15} & \cellcolor[HTML]{63BE7B}100\% & 00:16 & 00:16 & \cellcolor[HTML]{63BE7B}100\% \\ \hline
Modbus & DI2 & \textbf{00:09} & \textbf{00:09} & \cellcolor[HTML]{63BE7B}100\% & 00:55 & 01:10 & \cellcolor[HTML]{73C589}90\% \\
 & DI3 & \textbf{00:09} & \missing & \cellcolor[HTML]{BFE4CB}40\% & 01:19 & \missing & \cellcolor[HTML]{FCFCFF}0\% \\
\end{tabular}
}
\caption{Median bug survival times---both \textbf{R}eached and \textbf{T}riggered---over a 24-hour period across 10 trials for the \namebenchdma{} set, reported in HH:MM. “Hit” denotes the percentage of trials in which the bug was successfully triggered. The best-performing times are in bold. Table and statistics were generated using the Analysis Bench (see Fig.~\ref{figure:overview}) while other time-to-bug results are deferred to~\cite{FirmReBuggerGit}.}
\label{appendix:firmbenchdma01}
\end{table}


\noindent\textbf{DMA.~}Consider bug H13, shown in Figure~\ref{figure:survivial}, the challenge version (\bin{Contiki-RouterX}) reintroduces a patch restoring DMA  handling in the RF packet reception path. In the original binary, both \hoedur{} and \multifuzz{} could reliably trigger the bug. However, the DMA patch fundamentally alters how input data is processed: received radio packets are now transferred directly into memory via DMA, bypassing the traditional, byte-wise CPU-driven reads. Consequently, the fuzzers are blind to the data path that reaches the vulnerable code, preventing these fuzzers from triggering the bug, as shown in Figure~\ref{figure:survivial}.

For the DMA-focused fuzzers \dice{} and \gdma, our findings---summarized in Figure~\ref{figure:upsetX}~(b)---indicate that \gdma~\cite{scharnowskigdma} outperforms \dice~\cite{dice2021} on our \namebenchdma{} targets. Table~\ref{appendix:firmbenchdma01} presents the median reached and trigger times across bugs for both fuzzers. \gdma~\cite{scharnowskigdma} consistently achieves faster times, particularly on the \bin{Modbus} target (e.g., DI2 and DI3), where it reaches and triggers the bugs in under 10 minutes compared to \dice{'s}~\cite{dice2021} 55-79 minutes. While trigger rates are comparable on some bugs, the reduced time-to-bug demonstrates that \gdma~\cite{scharnowskigdma} exercises DMA-related code paths more efficiently.

\begin{mdframed}[backgroundcolor=blue!10,rightline=false,leftline=false,topline=false,bottomline=false,roundcorner=2mm] 
    \textbf{Key Takeaway}: Fuzz data injection into DMA remain challenging---with only limited developments~\cite{dice2021, scharnowskigdma, Dymafuzz} pursuing fully automated methods---because DMA controller implementations vary significantly by manufacturer or product lines. But, DMA inputs can guard critical code paths and are prevalent in firmware~\cite{scharnowskigdma} (10 binaries in our \namebenchdma{} set include DMA as a roadblock).
\end{mdframed}

\noindent
\textbf{Execution Delays and Input Bloating.~}
The roadblock in the \bin{CNC} and \bin{CNCX} binaries with bug FW18 is through a delay function that is reinstated in the challenge version. This delay function uses a busy-wait loop, forcing fuzzers to wait for the delay to elapse. This additional timing constraint causes all fuzzers to take longer to trigger the bug in the challenge binary as seen in Figure~\ref{figure:survivial}.

Meanwhile, in the \bin{Contiki-NG-Shell} and \bin{Contiki-NG-ShellX} binaries with bug S06 (see Figure~\ref{figure:survivial}, \bin{Contiki-NG-ShellX}) reintroduces the \texttt{fade} function, that executes thousands of loop iterations and MMIO accesses. This function requires the fuzzer to provide sufficiently long inputs in order to progress through the entire loop and any triggered interrupt handlers before reaching the buggy code path. As a result, only fuzzers with explicit length-extension strategies---as seen with \multifuzz{}---are able to consistently trigger the bug in the challenge binary. Other fuzzers frequently stall within the loop.

The \bin{BetaFlight} binary with bug FRB52, as illustrated in Figure~\ref{figure:survivial}, incorporates all of the identified roadblocks, making it exceptionally challenging to uncover the bug. Consequently, all existing fuzzer are unable to trigger the bug.
Discovery of the bug requires manual analysis, patching, and providing far more than 24 hours of fuzzing time.

\begin{mdframed}[backgroundcolor=blue!10,rightline=false,leftline=false,topline=false,bottomline=false,roundcorner=2mm] 
    \textbf{Key Takeaway}: The use of functions that, for example, execute excessive loop iterations---particularly for accessing MMIO---hinder a fuzzer from making progress to discover bugs. These functions can reduce test throughput, or lead to bloated inputs making mutations less effective.
\end{mdframed}

\noindent
\textbf{False Postive Bugs.~}Although false positive bugs are undesirable, they can demonstrate a fuzzer's ability to reach deep execution paths~\cite{scharnowski2022fuzzware}. Therefore, we retain false positive bugs in \namebench{}, clearly marking them as such. Notably, some false positives are specific to certain fuzzers, which should be considered when interpreting results. We discuss false positives further in Appendix~\ref{appendix:false-positives}.

\subsection{Extending With Community Contributions}\label{sec:extensibility}
Importantly, we envision a continuing expansion of \namebench{} through submission of Ravens, preventing overfitting and allowing for continued effective evaluation as fuzzers develop. Ravens submitted with a thorough root cause analysis of new bugs (similar to bug disclosures), where the community validates the Raven accurately encapsulates the bug’s semantics, can be merged into \namebench{}. In addition to manual reviews, accurate encapsulation can be validated by ensuring the Raven correctness as described in Section~\ref{sec:Raven-create-correct}.

We acknowledge the potential for community extensions to bias the benchmark targets or bugs in \namebench{} towards, for instance, a technique for a particular roadblock. While we have used some foresight and curated a DMA specific benchmark set for such an eventuality, the evolution of the field and the need to evaluate techniques may drive the creation of such subsets with bugs guarded by roadblocks. We envision the formation of a balanced binary set, that includes a balanced set of bugs from various categories and roadblocks as the field evolves. This would allow evaluation both at a wider level, and for techniques addressing specific bug types or roadblocks. The framework we developed is flexible to support these needs.
    
\subsection{Potential  \textsc{FirmBench} Dataset Bias}
The evaluated binaries are primarily sourced from prior works. This can potentially bias results towards these fuzzers. To minimize bias towards any one fuzzer, we curate our bug benchmark from bug reports across many different fuzzing works. As discussed in Section~\ref{sec:BechnmarkTargets}, we considered 10 SoTA works and add 3 new binary targets with 15 unique bugs whilst curating three target sets. Further, bug diversity, as discussed in Section~\ref{sec:diversity}, covering \uniqueCWECount{} unique CWEs prevents fuzzers targeting any one specific bug type from being unfairly advantaged and acts to mitigate bias. Importantly, \namebench{} is built to be extensible, facilitating the transparent introduction of new bugs over time to further mitigate biases and overfitting as discussed in Section~\ref{sec:extensibility}.

Additionally, \name{} consistently provides time-to-bug information for comparing fuzzers, a metric often not included in fuzzer evaluation \cite{feng2020p2im, zhou2022semu, dice2021, uEmu, farrelly2023ember, farrelly2023splits, chessermultifuzz, scharnowski2022fuzzware}. Time-to-bug measure allows works to demonstrate improved speed and consistency in triggering known bugs, beyond that observed for the fuzzer that originally discovered the bug.

\subsection{Triggered vs Detected}
We report bug count metrics based on the Triggered status because it can capture how well a fuzzer can explore code paths and program state of the space for bugs for a target. Whilst the detected status could be used, it can lead to under reporting bug finding capability as not all bugs triggered may result in a crash within a fuzzing campaign. It is important to acknowledge and appreciate that Triggered status based metrics may be influenced by differences in the emulators crash detection measures, or a fuzzer’s seed-saving strategy, since \name{} replays saved seeds and crashes. To avoid the possible impacts from differences, we support a LiveMode operation as we discuss in Section~\ref{livemode}.
Alternatively users of the benchmark can consider results based on both Detected and Triggered status when reporting bug counts. 

\subsection{FirmReBugger \textit{Live-Mode} Debugging}
\label{livemode}
Importantly, to enable interactive investigation of bugs, we implement a \textit{Live-Mode} of operation that can be employed during fuzzing by exploiting our Ravens and the provided API for emulators. We allow Ravens to be active or inactive. Active Ravens function as safeguards to identify if a triggering condition for a bug during execution is satisfied, and immediately crash. This allows a fuzzer to search for new vulnerabilities without the impact of previously known bugs, such as from bug exploits, making \name{} a flexible tool for targeted debugging and analysis.

\subsection{Related Work}
The need for methods to fairly assess fuzzers has led to the creation of benchmarking frameworks, \textit{outside} of the firmware domain, notably with Klees et al.\cite{klees2018evaluating}, LAVA\cite{dolan2016lava}, Magma~\cite{hazimeh2020magma}, UNIFUZZ~\cite{li2021unifuzz}, FuzzBench~\cite{metzman2021fuzzbench}, FixReverter~\cite{zhang2022fixreverter}, and FuzzProBench~\cite{profuzzbench}. These suites aim to provide curated collections of  binaries with realistic bugs to directly evaluate \textit{bug-finding} capabilities through bug counts and time-to-bug metrics. Prior work to design benchmarks considered dynamic injection of synthetic bugs, as in LAVA~\cite{dolan2016lava}, manually reverting bug patches as in Magma~\cite{hazimeh2020magma} and the DARPA Cyber Grand Challenge~\cite{darpa}, and automatic patch reversion, as in FixReverter~\cite{zhang2022fixreverter}. Notably, representativeness of synthetic bug injections remains debated~\cite{hazimeh2020magma, dolan2016lava}, further FixReverter depends on static analysis to determine bug reachability---a difficult task in firmware characterized by complex hardware interactions. Ours, similar to Magma, considers real-world targets and bugs but address the leaky oracle problem~\cite{hazimeh2020magma} by proposing a means for automatic triaging during replay of fuzz inputs without modifying the targets. 

\section{Conclusion}
\name{} is a bug-based benchmark designed to address the distinct challenges of monolithic firmware fuzzing. With carefully constructed real-world binaries and a standardized and automated evaluation workflow, we demonstrate and offer the community a reliable method for rigorous, reproducible assessment of monolithic firmware fuzzing techniques. Looking ahead, we envision our benchmark evolving alongside the field, accommodating new insights and techniques as they emerge. Ultimately, we hope \name{} will help speed-up developments and streamline  evaluation.

\section*{Acknowledgements}
The work was supported by the Cyber Security Research Centre Limited whose activities were partially funded by the Australian Government's Cooperative Research Centres Programme.
\appendix

\section{Ethical Considerations}
This work presents a benchmark for evaluating firmware fuzzing tools. Our intent is to support the security community in developing and evaluating effective fuzzing techniques, thereby improving the security of embedded systems to improve the safety and security of end-users of systems and devices with embedded systems.

In considering the datasets for the benchmark, we decided to employ firmware images  either open-source or publicly released, and no proprietary or personal data is included. We examined past state-of-the-art studies with public scrutiny of their work to source these firmware images. Consequently, the sourced software bugs were known and disclosed to project owners and maintainers.

We acknowledge that the benchmark and tools developed in this work have dual-use potential: while they are designed to support security researchers and improve the security of embedded systems, the tool developed from or tested with \name{} to improve their effectiveness could also be misused to identify vulnerabilities for malicious purposes by adversarial actors. These risks are partially reduced by our decision to employ a dataset of open-source or publicly available firmware and software bugs disclosed responsibly for active projects with support from maintainers.

Further, we also acknowledge that fuzzing the binaries used in our benchmark may lead to the discovery of previously unknown vulnerabilities. When such cases arise, these should be disclosed following responsible disclosure practices; reporting these findings directly to the relevant authors or maintainers prior to any public release or inclusion in the benchmark and providing sufficient time for maintainers to address the bug and release patches.

Ultimately, we believe, along with the precautions and commitment to following responsible disclosure practices for software bugs, the benefits of advancing defensive research through open science and strengthening system security outweigh the risks.

\section{Open Science}
To promote transparency and foster further advancements in firmware fuzzing research, we have open-sourced all code and materials associated with this work. These artifacts are publicly available on \href{https://github.com/FirmReBugger/FirmReBugger}{https://github.com/FirmReBugger/FirmReBugger}. 

\bibliographystyle{plain}
\bibliography{bibliography.bib}

\section{False Positive Bugs}\label{appendix:false-positives}
False positives are common when emulating embedded systems due to the close interaction between firmware and hardware and the inherent inaccuracy of emulation. A false positive occurs when a fuzzer reports a crash that arises from an emulated state that cannot occur on real hardware. A common source of these erroneous states come from interrupt injection. Emulators often fire interrupts pseudo-randomly, producing states that are impossible on actual devices.

False positives impact fuzzing campaigns in several ways. Crashing seeds influence the fuzzer's feedback and seed scheduling, while the increased number of crashes increases triage effort. Although false positive bugs are undesirable, they still demonstrate a fuzzer's ability to reach deep execution paths and find bugs. Furthermore, as highlighted by \fuzzware{}~\cite{scharnowski2022fuzzware}, these false positives often uncover hidden assumptions at the hardware/firmware boundary. Assumptions that may not always hold across different hardware contexts. This makes such findings valuable, as they can reveal potential vulnerabilities if the firmware is deployed on different hardware. Therefore, we retain false positives in our results, clearly marking them as such. Notably, some false positives are specific to certain fuzzers, which should be considered when interpreting the results.

\section{Benchmark Evaluation and Setup}
\label{appendix:evaluation-setup}
\label{appendix:benchmark-settings}
\noindent
\textbf{\namebench{}.~}For our baseline evaluations, we select seven recent state-of-the-art firmware fuzzers: \ember{} \cite{farrelly2023ember}, \fuzzware{} \cite{scharnowski2022fuzzware}, \fuzzwareicicle{} \cite{chesser2023icicle}, \semu{} \cite{zhou2022semu}, \splits{} \cite{farrelly2023splits}, \hoedur{} \cite{scharnowski2023hoedur}, and \multifuzz{} \cite{chessermultifuzz}. Where applicable, we use \fuzzware’s~\cite{scharnowski2022fuzzware} configuration files as the definition for the target’s memory map. 
For~\ember~\cite{farrelly2023ember} and \semu~\cite{zhou2022semu} we generate an equivalent configuration in a supported format. However, as \semu~\cite{zhou2022semu} requires manual, platform-specific set up for new targets, its evaluation is limited to the subset of binaries with support already available for target of boards: F103, F429, K64 and SAM3X. Additionally, due to some instructions not being implemented within the version of Unicorn\cite{quynh2015unicorn} used by \fuzzware{} and \splits, we exclude the \bin{Oresat-Control}---see Table~\ref{table:summary_binaries}---binary from the evaluation of these fuzzers.

\vspace{1mm}
\noindent
\textbf{\namebenchdma{}.~}Two recent fully-automated works~\cite{scharnowskigdma,dice2021} focus on the extending existing firmware fuzzers to support DMA using information gathered at runtime. To allow for comparison of these works, we include two bug-containing DMA binaries compatible with both \gdma{} and \dice{}.

\vspace{1mm}
\noindent
\textbf{\namebenchX{}.~}For evaluations with \namebenchX{}, we consider the fuzzers used with \namebench{}, except \semu{}~\cite{zhou2022semu} due to \semu{}'s dependence on additional manually configured components, and the limited exploration established in \namebench{}, we exclude \semu{} from the \namebenchX{} tests.

\onecolumn

\renewcommand{\thetable}{A\arabic{table}}
\renewcommand{\thefigure}{A\arabic{figure}}
\setcounter{table}{0}
\setcounter{figure}{0}
\section{Benchmark Bug Descriptions}\label{appendix:bug-descriptions}
\begin{table}[h]
\centering
\caption{Summary of fuzzing roadblocks identified in each binary in the \namebench{} benchmark. Symbols indicate subset inclusion: $\dag$~\namebenchX{}, $\ddag$~\namebenchdma{}, and $\ast$~present across all benchmark subsets. We use the presence of calls to \texttt{strcmp}, or \texttt{strncmp} as a proxy for the presence of magic values. Complex peripherals are determined based on access to SPI or USB interfaces. DMA is ticked for any binaries that perform transactions with the DMA controller. Delay \& Bloat is ticked if functions such as \texttt{fade} or \texttt{delay} are called that bloat inputs or intentionally delay further execution by executing loops without any other purpose}
\label{table:summary_binaries}
\resizebox{\textwidth}{!}{%
\begin{tabular}{l|l|l|l|c|c|c|c|l}
\hline
REF                                                       & Binary                                    & Blocks & OS/Sys lib.               & \multicolumn{1}{l|}{Magic Value} & \multicolumn{1}{l|}{Cplx. Peripherals} & \multicolumn{1}{l|}{DMA} & \multicolumn{1}{l|}{Delay \& Bloat} & BugID (CWE)                                                                                                                                                                                                                                                                                                                                                                                                                                                                                             \\ \hline
\multirow{5}{*}{\ptwoim~\cite{feng2020p2im}}              & CNC \dag                                  & 3615   & Bare metal                & \cross                           & \tick                                  & \cross                   & \tick                               & \textbf{FP\_E02}, \textbf{FW11}(CWE-121), \textbf{FW18}(CWE-1286)                                                                                                                                                                                                                                                                                                                                                                                                                                       \\ \cline{2-9} 
                                                          & Console \dag                              & 2225   & RIOT~\cite{riotos_cpus}   & \tick                            & \cross                                 & \cross                   & \tick                               & \textbf{I01}(CWE-1284)                                                                                                                                                                                                                                                                                                                                                                                                                                                                                  \\ \cline{2-9} 
                                                          & Gateway \dag                              & 4922   & Arduino~\cite{arduino}    & \tick                            & \tick                                  & \cross                   & \tick                               & \begin{tabular}[c]{@{}l@{}}\textbf{E01}(CWE-252), \textbf{FP\_FRB01}, \textbf{FP\_FRB02}, \\ \textbf{FP\_FRB03}, \textbf{FP\_FRB04}, \textbf{FP\_FRB10}, \\ \textbf{FP\_FW21}, \textbf{FP\_FW22}, \textbf{FW12}(CWE-787), \\ \textbf{FW23}(CWE-825), \textbf{MF01}(CWE-1284)\end{tabular}                                                                                                                                                                                                               \\ \cline{2-9} 
                                                          & PLC \dag                                  & 2304   & Arduino~\cite{arduino}    & \cross                           & \tick                                  & \cross                   & \tick                               & \begin{tabular}[c]{@{}l@{}}\textbf{FP\_FW25}, \textbf{FW14}(CWE-1285), \textbf{FW15}(CWE-1285), \\ \textbf{FW16}(CWE-1285), \textbf{FW17}(CWE-1285)\end{tabular}                                                                                                                                                                                                                                                                                                                                        \\ \cline{2-9} 
                                                          & Soldering\_Iron                           & 3657   & FreeRTOS~\cite{freertos}  & \cross                           & \tick                                  & \cross                   & \tick                               & \textbf{FP\_FRB05}, \textbf{FP\_I02}, \textbf{FP\_MF02}, \textbf{FW19}(CWE-825), \textbf{H01}(CWE-825)                                                                                                                                                                                                                                                                                                                                                                                                  \\ \hline
\textsc{PRETENDER}\cite{gustafson2019Pretender}           & RF\_Door\_Lock                            & 3321   & Mbed~\cite{mbedos}        & \tick                            & \tick                                  & \cross                   & \tick                               & \textbf{FRB09}(CWE-120), \textbf{FW29}(CWE-121), \textbf{FW38}(CWE-674)                                                                                                                                                                                                                                                                                                                                                                                                                                 \\ \hline
\multirow{5}{*}{\uemu~\cite{uEmu}}                        & 3DPrinter \dag                            & 8046   & Arduino~\cite{arduino}    & \tick                            & \tick                                  & \cross                   & \tick                               & \textbf{FP\_FW39}                                                                                                                                                                                                                                                                                                                                                                                                                                                                                       \\ \cline{2-9} 
                                                          & GPSTracker \dag                           & 4195   & Arduino~\cite{arduino}    & \tick                            & \tick                                  & \cross                   & \tick                               & \begin{tabular}[c]{@{}l@{}}\textbf{FW30}(CWE-121), \textbf{FW31}(CWE-690), \textbf{MF02}(CWE-690), \textbf{MF03}(CWE-690), \\ \textbf{S01}(CWE-690), \textbf{S02}(CWE-690)\end{tabular}                                                                                                                                                                                                                                                                                                                 \\ \cline{2-9} 
                                                          & utasker~\cite{utasker}\_USB               & 3492   & utasker~\cite{utasker}    & \tick                            & \tick                                  & \cross                   & \tick                               & \begin{tabular}[c]{@{}l@{}}\textbf{FP\_FRB06}, \textbf{FP\_FRB07}, \textbf{FP\_FRB08}, \textbf{FP\_FW27}, \textbf{FP\_FW45}, \\ \textbf{FP\_MF05}, \textbf{MF04}(CWE-1285), \textbf{S04}(CWE-123)\end{tabular}                                                                                                                                                                                                                                                                                          \\ \cline{2-9} 
                                                          & utasker~\cite{utasker}\_MODBUS            & 3781   & utasker~\cite{utasker}    & \tick                            & \tick                                  & \cross                   & \tick                               & \textbf{FP\_FW40}, \textbf{S03}(CWE787)                                                                                                                                                                                                                                                                                                                                                                                                                                                                 \\ \cline{2-9} 
                                                          & Zephyr~\cite{zephyr}\_SocketCan           & 5944   & Zephyr~\cite{zephyr}      & \tick                            & \cross                                 & \cross                   & \tick                               & \begin{tabular}[c]{@{}l@{}}\textbf{E03}(CWE-120), \textbf{FP\_FRB11}, \textbf{FP\_FRB12}, \textbf{FP\_FW44}, \\ \textbf{FW43}(CWE-457), \textbf{MF06}(CWE-685), \textbf{MF07}(CWE-685), \\ \textbf{MF08}(CWE-685), \textbf{MF10}(CWE-843), \textbf{MF11}(CWE-843), \textbf{MF12}(CWE-843), \\ \textbf{MF13}(CWE-843), \textbf{MF14}(CWE-843), \textbf{MF15}(CWE-843), \textbf{MF16}(CWE-843), \\ \textbf{FP\_MF09}, \textbf{S05}(CWE-822)\end{tabular}                                                  \\ \hline
\textsc{HALucinator}\cite{clements2020halucinator}        & 6LoWPAN\_Receiver                         & 6978   & Contiki~\cite{contiki-ng} & \tick                            & \tick                                  & \cross                   & \tick                               & \textbf{FP\_E04}, \textbf{FP\_MF18}, \textbf{FW36}(CWE-457), \textbf{MF17}(CWE-1285)                                                                                                                                                                                                                                                                                                                                                                                                                    \\ \hline
\splits~\cite{farrelly2023splits}                         & Contiki~\cite{contiki-ng}\_NG\_Shell \dag & 4777   & Contiki~\cite{contiki-ng} & \tick                            & \tick                                  & \cross                   & \tick                               & \textbf{S06}(CWE-685)                                                                                                                                                                                                                                                                                                                                                                                                                                                                                   \\ \hline
\multirow{2}{*}{\dice~\cite{dice2021}}                        & MIDI \ddag                                & 810    & Bare metal                & \cross                           & \tick                                  & \tick                    & \cross                              & \textbf{DI4}(CWE590), \textbf{DI5}(CWE590)                                                                                                                                                                                                                                                                                                                                                                                                                                                              \\ \cline{2-9} 
                                                          & MODBUS \ddag                              & 811    & FreeRTOS~\cite{freertos}  & \cross                           & \tick                                  & \tick                    & \tick                               &  \textbf{DI2}(CWE193), \textbf{DI3}(CWE193)                                                                                                                                                                                                                                                                                                                                                                                                                                        \\ \hline
\multirow{8}{*}{\fuzzware~\cite{scharnowski2022fuzzware}} & Contiki~\cite{contiki-ng}-hello-4-4*      & 3960   & Contiki~\cite{contiki-ng} & \tick                            & \cross                                 & \tick                    & \tick                               & \begin{tabular}[c]{@{}l@{}}\textbf{H03}(CWE-131), \textbf{H04}(CWE-674), \textbf{FW58}(CWE-120), \textbf{H06}(CWE-787), \\ \textbf{H07}(CWE-787), \textbf{H08}(CWE-787), \textbf{H09}(CWE-1288), \textbf{H10}(CWE-1288)\end{tabular}                                                                                                                                                                                                                                                                    \\ \cline{2-9} 
                                                          & Contiki~\cite{contiki-ng}-6lowpan*        & 3082   & Contiki~\cite{contiki-ng} & \tick                            & \cross                                 & \tick                    & \tick                               & \textbf{HAL02}(CWE-787), \textbf{HAL01}(CWE-191)                                                                                                                                                                                                                                                                                                                                                                                                                                                        \\ \cline{2-9} 
                                                          & Contiki~\cite{contiki-ng}-snmp*           & 3039   & Contiki~\cite{contiki-ng} & \tick                            & \cross                                 & \tick                    & \tick                               & \textbf{FW59}(CWE-120), \textbf{H15}(CWE-125), \textbf{H16}(CWE-770), \textbf{H17}(CWE-787)                                                                                                                                                                                                                                                                                                                                                                                                             \\ \cline{2-9} 
                                                          & Zephyr~\cite{zephyr}-3330 \dag            & 6867   & Zephyr~\cite{zephyr}      & \tick                            & \tick                                  & \cross                   & \tick                               & \textbf{FW55}(CWE-787)                                                                                                                                                                                                                                                                                                                                                                                                                                                                                  \\ \cline{2-9} 
                                                          & Zephyr~\cite{zephyr}-bt \dag              & 4907   & Zephyr~\cite{zephyr}      & \tick                            & \tick                                  & \cross                   & \tick                               & \textbf{FW48}(CWE476), \textbf{FW47}(CWE787)                                                                                                                                                                                                                                                                                                                                                                                                                                                            \\ \cline{2-9} 
                                                          & Zephyr~\cite{zephyr}-nrf \dag             & 4938   & Zephyr~\cite{zephyr}      & \tick                            & \tick                                  & \cross                   & \tick                               & \begin{tabular}[c]{@{}l@{}}\textbf{FW54}(CWE-665), \textbf{H34}(CWE-476), \textbf{H35}(CWE-704), \textbf{H33}(CWE-665),\\ \textbf{H36}(CWE-697), \textbf{H37}(CWE-787), \textbf{H38}(CWE-703), \textbf{H39}(CWE-703), \\ \textbf{H40}(CWE-703), \textbf{H42}(CWE-665), \textbf{H43}(CWE-476), \\ \textbf{H44}(CWE-362), \textbf{H45}(CWE-665), \textbf{H46}(CWE-457), \textbf{H47}(CWE-476), \\ \textbf{H48}(CWE-476), \textbf{H49}(CWE-703), \textbf{H50}(CWE-476), \textbf{H51}(CWE-125)\end{tabular} \\ \cline{2-9} 
                                                          & Zephyr~\cite{zephyr}-sam4s \dag           & 6954   & Zephyr~\cite{zephyr}      & \tick                            & \tick                                  & \cross                   & \tick                               & \begin{tabular}[c]{@{}l@{}}\textbf{FP\_FRB16}, \textbf{FP\_FRB17}, \textbf{H52}(CWE-476), \textbf{FW53}(CWE-191), \\ \textbf{FW50}(CWE-476), \textbf{FW49}(CWE-476), \textbf{H51}(CWE-191), \textbf{H62}(CWE-191)\end{tabular}                                                                                                                                                                                                                                                                          \\ \cline{2-9} 
                                                          & Zephyr~\cite{zephyr}-sampro \dag          & 7176   & Zephyr~\cite{zephyr}      & \tick                            & \tick                                  & \cross                   & \tick                               & \begin{tabular}[c]{@{}l@{}}\textbf{FP\_FRB18}, \textbf{FP\_FRB19}, \textbf{FP\_FRB30}, \textbf{FW46}(CWE-787), \\ \textbf{FW50}(CWE-476), \textbf{FW52}(CWE-476), \textbf{FW51}(CWE-191), \textbf{FW53}(191), \\ \textbf{H59}(CWE-476), \textbf{H60}(CWE-476), \textbf{H61}(CWE-476)\end{tabular}                                                                                                                                                                                                       \\ \hline
\multirow{5}{*}{\hoedur~\cite{scharnowski2023hoedur}}     & Contiki~\cite{contiki-ng}-hello-4-8*      & 3988   & Contiki~\cite{contiki-ng} & \tick                            & \cross                                 & \tick                    & \tick                               & \begin{tabular}[c]{@{}l@{}}\textbf{FRB20}(CWE-125), \textbf{H03}(CWE-131), \textbf{H04}(CWE-674), \textbf{FW58}(CWE-120), \\ \textbf{H06}(CWE-787), \textbf{H07}(CWE-787), \textbf{H08}(CWE-787), \textbf{H09}(CWE-1288), \\ \textbf{H10}(CWE-1288), \textbf{H11}(CWE-125), \textbf{H12}(CWE-922)\end{tabular}                                                                                                                                                                                          \\ \cline{2-9} 
                                                          & Contiki~\cite{contiki-ng}-router*         & 4198   & Contiki~\cite{contiki-ng} & \tick                            & \tick                                  & \tick                    & \tick                               & \textbf{H13}(CWE-476)                                                                                                                                                                                                                                                                                                                                                                                                                                                                                   \\ \cline{2-9} 
                                                          & loramac \dag                              & 5264   & Zephyr~\cite{zephyr}      & \tick                            & \tick                                  & \cross                   & \tick                               & \begin{tabular}[c]{@{}l@{}}\textbf{FP\_FRB25}, \textbf{FP\_FRB26}, \textbf{H18}(CWE-193), \\ \textbf{FP\_FRB58}\end{tabular}                                                                                                                                                                                                                                                                                                                                                                            \\ \cline{2-9} 
                                                          & gnrc\_networking                          & 11912  & RIOT~\cite{riotos_cpus}   & \tick                            & \cross                                 & \cross                   & \tick                               & \begin{tabular}[c]{@{}l@{}}\textbf{FP\_FRB23}, \textbf{FP\_FRB31}, \textbf{FP\_FRB32}, \textbf{FP\_FRB35}, \\ \textbf{FP\_FRB36}, \textbf{FP\_FRB37}, \textbf{H19}(CWE-191), \textbf{H20}(CWE-476), \\ \textbf{H21}(CWE-787), \textbf{H22}(CWE-191), \textbf{H23}(CWE-191), \textbf{H24}(CWE-476), \\ \textbf{H25}(CWE-787), \textbf{H26}(CWE-252), \textbf{H27}(CWE-824)\end{tabular}                                                                                                                  \\ \cline{2-9} 
                                                          & Zephyr~\cite{zephyr}-f429zi \dag          & 5501   & Zephyr~\cite{zephyr}      & \tick                            & \tick                                  & \cross                   & \tick                               & \textbf{H31}(CWE-415)                                                                                                                                                                                                                                                                                                                                                                                                                                                                                   \\ \hline
\multirow{2}{*}{\multifuzz~\cite{chessermultifuzz}}       & RIOT~\cite{riotos_cpus}\_CCN\_LITE        & 12675  & RIOT~\cite{riotos_cpus}   & \tick                            & \cross                                 & \cross                   & \tick                               & \begin{tabular}[c]{@{}l@{}}\textbf{FP\_MF22}, \textbf{MF19}(CWE-362), \textbf{MF20}(CWE-664), \textbf{MF21}(CWE-672), \\ \textbf{MF22}(CWE-177)\end{tabular}                                                                                                                                                                                                                                                                                                                                            \\ \cline{2-9} 
                                                          & RIOT~\cite{riotos_cpus}\_GNRC             & 6449   & RIOT~\cite{riotos_cpus}   & \tick                            & \cross                                 & \cross                   & \tick                               & \textbf{MF23}(CWE-191)                                                                                                                                                                                                                                                                                                                                                                                                                                                                                  \\ \hline
\multirow{3}{*}{\textsc{FirmReBugger}}                      & Hoverboard*                               & 4440   & ChibiOS~\cite{chibios}    & \tick                            & \tick                                  & \tick                    & \tick                               & \textbf{FP\_FRB52}, \textbf{FRB38}(CWE-252), \textbf{FRB39}(CWE-457), \textbf{FRB46}(CWE-822)                                                                                                                                                                                                                                                                                                                                                                                                           \\ \cline{2-9} 
                                                          & Oresat-Control*                           & 20504  & ChibiOS~\cite{chibios}    & \tick                            & \tick                                  & \tick                    & \tick                               & \begin{tabular}[c]{@{}l@{}}\textbf{FP\_FRB53}, \textbf{FP\_FRB54}, \textbf{FP\_FRB55}, \textbf{FRB48}(CWE-685), \\ \textbf{FRB49}(CWE-685), \textbf{FRB50}(CWE-685), \textbf{FRB51}(CWE-685)\end{tabular}                                                                                                                                                                                                                                                                                               \\ \cline{2-9} 
                                                          & BetaFlight*                               & 26319  & Bare metal                & \tick                            & \tick                                  & \tick                    & \tick                               & \textbf{FRB52}(CWE-685), \textbf{FRB54}(CWE-685), \textbf{FRB55}(CWE-252), \textbf{FRB56}(CWE-457)                                                                                                                                                                                                                                                                                                                                                                                                     
\end{tabular}
}
\end{table}
\twocolumn
\end{document}